\documentclass[preprint,showpacs,preprintnumbers,amsmath,amssymb,showkeys,nofootinbib]{revtex4}
   
\usepackage{graphicx}
\usepackage{dcolumn}
\usepackage{bm}
\usepackage{braket}
 
\begin{document} 

\noindent

\preprint{}

\title{Quantum mechanics is a calculus for estimation under epistemic restriction}

\author{Agung Budiyono} 
\email{agungbymlati@gmail.com}
\affiliation{Research Center for Nanoscience and Nanotechnology, Bandung Institute of Technology, Bandung, 40132, Indonesia}
\affiliation{Edelstein Center, Hebrew University of Jerusalem, Jerusalem, 91904 Israel} 
\affiliation{Department of Engineering Physics, Bandung Institute of Technology, Bandung, 40132, Indonesia}
\affiliation{Kubus Computing and Research, Juwana, Pati, 59185 Indonesia}
 
\date{\today}
   
\begin{abstract}   

Consider a statistical model with an epistemic restriction such that, unlike in classical mechanics, the allowed distribution of positions is fundamentally restricted by the form of an underlying momentum field. Assume an agent (observer) who wishes to estimate the momentum field given information on the conjugate positions. We discuss a classically consistent, weakly unbiased, best estimation of the momentum field minimizing the mean squared error, based on which the abstract mathematical rules of quantum mechanics can be derived. The results suggest that quantum wave function is not an objective agent-independent attribute of reality, but represents the agent's best estimation of the momentum, given the positions, under epistemic restriction. Quantum uncertainty and complementarity between momentum and position find their epistemic origin from the trade-off between the mean squared errors of simultaneous estimations of momentum field and mean position, with the Gaussian wave function represents the simultaneous efficient estimations, achieving the Cram\'er-Rao bounds of the associated mean squared errors. We then argue that unitary time evolution and wave function collapse in measurement are normative rules for an agent to update her/his estimation given information on the experimental settings.  
     
\end{abstract}     
 
\maketitle  

\section{Introduction}       

Quantum mechanics is without doubt an incomparably successful physical theory, and it plays a decisive role in our attempt to comprehend Nature. It is therefore remarkable that, after almost a century since its inception, the discussion on its meaning is as hot as ever. Central to this debate is on the meaning of pure quantum state or wave function, i.e., whether wave function is an objective physical thing independent of measurement \cite{Bohm pilot-wave theory,Everett many worlds,Wheeler many worlds,deWitt many worlds,GRW theory}, or, it is merely a convenient mathematical tool which represents the agent's subjective knowledge or information about the system \cite{Bohr Q epistemic,Ballentine statistical interpretation,Griffiths consistent histories,Omnes consistent histories,Gell-Mann and Hartle decoherent histories,Rovelli relational QM,Zeilinger axioms,Caves informational wave function,Fuchs QBism}. A physical wave function, for example, can explain particle interference transparently. However, a physical wave which is defined in multidimensional configuration space (instead of in the ordinary three dimensional physical space), or suffers instantaneous collapse during measurement, is conceptually uncanny. By contrast, an informational wave function does not suffer from the latter two conceptual problems, but then particle interference becomes a mystery. The standard mathematical formalism of quantum mechanics does not provide a general unequivocal guide to unscramble parts of the theory which refer to objective reality from those which merely represent the agent's information about reality \cite{Jaynes ontic-epistemic,Harrigan ontic-epistemic}. A related fundamental problem is that the standard formalism does not offer a transparent quantum-classical boundary and correspondence \cite{Zurek quantum-classical correspondence,Emerson PhD thesis}.   

We might understand quantum mechanics better if we could derive it from simple axioms with transparent meaning, rather than the well-known, notoriously abstruse, formal mathematical axioms \cite{Wheeler: why,Popescu-Rohrlich axioms}. This reconstruction program is also pivotal to answer a tantalizing question why quantum mechanics has the specific form it does, expressed in an abstract complex Hilbert space, such that a slight modification of its axioms may lead to bizarre implications such as superluminal signalling \cite{Gisin nonlinearity - signaling,Polchinski nonlinearity - signaling,Lee nonHermitian - signaling} or violations of the second law of thermodynamics \cite{Peres nonlinearity violates 2nd law,Hanggi UR second law}. Various axiomatic derivations of quantum mechanics with different approaches and scopes have been suggested \cite{Rovelli relational QM,Caves informational wave function,Zeilinger axioms,Caves-Fuchs axioms,Hardy axioms,Fuchs axioms,Spekkens toy model with epistemic restriction,Barrett axioms,Dakic-Brukner axioms,Paterek axioms,Chiribella axioms,Masanes axioms,de la Torre axioms,Bartlett reconstruction of Gaussian QM with epistemic restriction,Spekkens quasi-quantization,Reginatto estimation scheme,Hall-Reginatto model,Markopolou-Smolin quantum from cosmos,Smolin quantum from cosmos,Nelson stochastic mechanics,Frieden model, deRaedt model,Agung-Daniel model,Goyal information geometry,Caticha quantum from MEP,Hohn quantum from question,Selby quantum from diagram}. In particular, significant efforts have been spent in the last decades to show that a remarkably large set of phenomena, traditionally regarded as specifically quantum, could in fact surprisingly be explained using classical statistical models by imposing some form of `epistemic restrictions' (statistical constraints), which attempt to capture an intuitive picture about quantum uncertainty \cite{Hardy epistricted model,Emerson PhD thesis,Spekkens toy model with epistemic restriction,Bartlett reconstruction of Gaussian QM with epistemic restriction,Spekkens quasi-quantization}. Such an approach is crucial to better understand the deep distinction between quantum and classical worlds, i.e. the fundamental origin of nonclassicality, which might find important practical applications to transparently identify the boundary between quantum and classical computations. Following this line of inquiry, In Ref. \cite{Agung-Daniel model}, we have proposed a statistical model wherein the abstract mathematical formalism of spin-less non-relativistic quantum mechanics is shown to emerge from a modification of classical mechanics, introducing a specific epistemic restriction, and an ontic extension in the form of a global-nonseparable random variable with a strength on the order of Planck constant, parameterizing the epistemic restriction. The epistemic restriction and ontic extension are argued to imply quantum uncertainty and entanglement. 

In the present work, we ask: what does the statistical model in Ref. \cite{Agung-Daniel model} tell us about the nature of wave function and thus the meaning of quantum mechanics? To answer this important interpretational question, we need to clarify the meaning of the specific epistemic restriction postulated in the model (see Eq. (\ref{fundamental epistemic restriction}) below), which gives rise to the mathematical expression of the wave function. We argue that the specific epistemic restriction in Ref. \cite{Agung-Daniel model} can be motivated from deeper and transparent informational constraints employed in the field of (classical) parameter estimation theory \cite{Papoulis and Pillai book on probability and statistics,Cover-Thomas book}. To do this, we first assume that unlike in classical mechanics, in microscopic world, there is a general epistemic restriction such that the allowed probability distribution of positions is fundamentally restricted, hence parameterized, by the form of an underlying momentum field. Consider an agent who wishes to estimate the momentum field, given information on the conjugate positions, under such an epistemic restriction. In this operational setting, we show that the specific epistemic restriction in \cite{Agung-Daniel model} can be seen naturally as describing a classically consistent, weakly unbiased, ``best'' estimation of the momentum field minimizing the mean squared (MS) error. 

The results suggest that wave function is not an objective agent-independent attribute of a reality, but is a mathematical encryption of an agent-dependent information about her/his preparation: it represents the agent's best estimation of the momentum, given the positions, in the presence of epistemic restriction. Quantum uncertainty and complementarity between momentum and position are then shown to originate from the trade-off between the MS errors of simultaneous estimations of momentum field and mean position, with the Gaussian wave function represents the simultaneous ``efficient'' estimations, attaining the Cram\'er-Rao bounds of the associated MS errors. Unitary Schr\"odinger equation arises naturally as a rule for updating the agent's estimation when she/he does not make any selection of trajectories so that her/his estimation must respect the conservation of trajectories and average energy. And, epistemic wave function collapse reflects the Bayesian updating of the agent's estimation due to a selection of certain subset of trajectories compatible with the measurement outcomes. Within the epistemic reconstruction and interpretation, the Wigner's friend paradox \cite{Wigner's paradox} is explained in term of two different estimations by two agents having two incompatible information. 

Let us mention that different approaches to derive the Schr\"odinger equation invoking the ideas of estimation were reported in Refs. \cite{Frieden model,Reginatto estimation scheme}. Unlike our approach which is based on estimation of momentum given information on the conjugate position under a fundamental assumption of epistemic restriction, in these latter approaches, they start from estimation of position, and minimize the associated Fisher information, under some statistical constraints. An approach where Schr\"odinger equation is obtained from statistical inference employing the maximum entropy principle to the entropy of a newly added variable is proposed in Ref. \cite{Caticha quantum from MEP}, and, in Ref. \cite{deRaedt model} the authors argue that the Schr\"odinger equation can be derived from logical inference applied to robust experiments. Next, in Ref. \cite{Hall-Reginatto model} the authors use the exact uncertainty relation defined in \cite{Hall exact uncertainty and best estimation}, to infer the form of the Lagrangian associated with an ensemble and use variational principle to obtain the Schr\"odinger equation. See also Ref. \cite{Deutch Born's rule from decision} for a derivation of Born's rule based on decision theory, and Ref. \cite{Caves informational wave function} for the interpretation of Gleason theorem for the derivation of Born's rule as reflecting Bayesian reasoning based on Dutch-book argument, leading to a Bayesian interpretation of quantum mechanics \cite{Fuchs QBism}.  

\section{Estimation of momentum given information on conjugate positions under epistemic restriction}

For simplicity, we consider first a general classical system of one spatial dimension $q$ with the conjugate momentum $p$. Let $t$ denotes time. We recall that within the Hamilton-Jacobi formalism of classical mechanics, the momentum field arising in a preparation can in general be expressed as:
\begin{eqnarray}
\tilde{p}_{\rm C}(q,t)=\partial_qS_{\rm C}(q,t),
\label{Hamilton-Jacobi condition}
\end{eqnarray} 
where $S_{\rm C}(q,t)$ is the Hamilton's principal function. (Here and below, we use $\tilde{p}$ to represent the functional form of momentum field, and $p$ is used to denote a specific value of momentum.) Moreover, for a system with a classical Hamiltonian $H(q,p)$, the Hamilton's principal function $S_{\rm C}(q,t)$ evolves with time following the Hamilton-Jacobi equation: $-\partial_tS_{\rm C}=H(q,p)=H(q,\partial_qS_{\rm C})$. Solving the Hamilton-Jacobi equation for $S_{\rm C}(q,t)$, a trajectory is singled out by selecting a position $q=q_0$ at an arbitrary time $t=t_0$, wherein the momentum along the trajectory is obtained by computing Eq. (\ref{Hamilton-Jacobi condition}). Hamilton-Jacobi formalism thus offers a geometrical description of an ensemble of trajectories, obtained by repeating the experiment many times varying $q=q_0$ at $t=t_0$, all following a momentum field  $\tilde{p}_{\rm C}(q,t)$ characterized by a single Hamilton's principal function $S_{\rm C}(q,t)$ \cite{Rund book: Hamilton-Jacobi formalism}.  

Let us discuss such an ensemble of classical trajectories and express the probability distribution that the system has a position $q$ at time $t$ as $\rho(q,t)$. It is then clear from the above Hamilton-Jacobi formalism that in classical mechanics, each trajectory in a given momentum field $\tilde{p}_{\rm C}(q)$ of Eq. (\ref{Hamilton-Jacobi condition}) can be assigned an arbitrary weight $\rho(q)$ (dependence on time is notationally omitted). For a simple illustration, consider a uniform momentum field $\tilde{p}_{\rm C}(q)=p_o$, where $p_o$ is spatially constant. Then, we can prepare any arbitrary probability distribution of position $\rho(q)$ by suitably weighting each point $q$ with $\rho(q)$. Hence, in classical mechanics, `the probability distribution of position $\rho(q)$ is in principle independent of, thus is not parameterized by, the underlying momentum field $\tilde{p}_{\rm C}(q)$'. We argue that this `epistemic freedom', i.e., the freedom to choose an arbitrary probability distribution of positions $\rho(q)$ independent of the underlying momentum field $\tilde{p}_{\rm C}(q)$, is a fundamental principle (implicitly) assumed in classical mechanics (analogous to the independency of time and space in pre-relativity physics). 

We assume that the above epistemic freedom is no longer fully granted in microscopic world (in an analogous way that absolute time, independent of spatial coordinate, is violated by fast moving objects in relativity theory). First, assume that in microscopic world, the momentum field arising in a preparation is fluctuating randomly. To this end, we introduce a global-nonseparable random variable $\xi$ the fluctuation of which induces the random fluctuation of the momentum field $\tilde{p}(q,t;\xi)$. We then assume that the ensemble of trajectories following the momentum field, obtained by repeating the experiment many times, suffer a general form of `epistemic restriction': namely, given a momentum field $\tilde{p}(q;\xi)$, unlike in classical mechanics discussed above, it is {\it no} longer possible to prepare an ensemble of trajectories with an arbitrary probability distribution of position, or, each trajectory in the momentum field can no longer be assigned an arbitrary weight \cite{Agung-Daniel model}. The allowed probability distributions of position therefore fundamentally depends on, thus is restricted or  parameterized by, the functional form of the underlying momentum field $\tilde{p}(q;\xi)$. To make explicit this dependency via statistical restriction (parameterization), we write the probability distribution of position as 
\begin{eqnarray}
\rho_{\tilde{p}}(q). \nonumber
\end{eqnarray} 

Now, suppose an agent makes a measurement of position and obtain a value $q$, e.g. via a selection of trajectories passing through $q$. (As emphasized by Bell, any measurement should be reducible to the measurement of position \cite{Bell book}. See also Ref. \cite{Popper book}.) Since $q$ is sampled from $\rho_{\tilde{p}}(q)$, it must somehow carry some information about the underlying momentum field $\tilde{p}(q;\xi)$ parameterizing $\rho_{\tilde{p}}(q)$. Our question is then: how can the agent use her/his information on $q$, in the most reasonable way, to estimate the form of the conjugate momentum field $\tilde{p}(q;\xi)$ parameterizing $\rho_{\tilde{p}}(q)$. This is a parameter estimation problem which has extensively been studied in (classical) statistical-information theory \cite{Papoulis and Pillai book on probability and statistics,Cover-Thomas book}. Note that in our problem, the parameter to be estimated, i.e., the random momentum field, is itself a function of the position used for estimation. Does Nature impose a fundamental informational restriction for the agent to estimate the momentum field arising in her/his preparation given information on the conjugate position? 

What are the requirements that a `good' estimator for the momentum field $\tilde{p}(q;\xi)$, based on information on $q$, should satisfy? First, it is imperative that the estimator has a consistent classical limit. Noting the fact that in macroscopic regime the momentum field takes the form of Eq. (\ref{Hamilton-Jacobi condition}), we assume that given $q$, the estimator for $\tilde{p}(q;\xi)$ has a general form $\partial_qS(q)$. Here, $S(q)$ is a real-valued function with the dimension of action, and we demand that in the classical limit we recover Eq. (\ref{Hamilton-Jacobi condition}). Next, we require the estimator $\partial_qS(q)$ to satisfy a `weak' unbiased condition so that for any value of $\xi$, the average of the estimator $\partial_qS(q)$ over $\rho_{\tilde{p}}(q)$, is equal to the average of the momentum field $\tilde{p}(q;\xi)$ to be estimated over $\rho_{\tilde{p}}(q)$, i.e.: $\int{\rm d}q~\partial_qS(q)\rho_{\tilde{p}}(q)=\int{\rm d}q~\tilde{p}(q;\xi)\rho_{\tilde{p}}(q)$. This requirement simply means that, for all $\xi$, the error of estimation $\tilde{p}(q;\xi)-S(q)$ is on average vanishing. As shown in the Appendix \ref{derivation of Cramer-Rao bound}, this reasonable requirement implies the Cram\'er-Rao inequality. Namely, for each $\xi$, the MS error of the estimation of momentum field $\tilde{p}(q;\xi)$ with the weakly unbiased estimator $\partial_qS(q)$ is bounded from below as
\begin{eqnarray}
\int{\rm d}q\big(\tilde{p}(q;\xi)-\partial_qS(q)\big)^2\rho_{\tilde{p}}(q)\ge\frac{1}{J_p},
\label{Cramer-Rao bound for momentum estimation}
\end{eqnarray}
where $J_p\doteq\int{\rm d}q\big(\partial_{\tilde{p}}\ln\rho_{\tilde{p}}(q)\big)^2\rho_{\tilde{p}}(q)$ is the Fisher information about the momentum field $\tilde{p}$ contained in $\rho_{\tilde{p}}(q)$. The weakly unbiased estimator $\partial_qS$ is called efficient if it saturates the Cram\'er-Rao inequality of Eq. (\ref{Cramer-Rao bound for momentum estimation}). 

Let us also assume that in the mathematical limit $\xi\rightarrow 0$, namely in the absence of the global random fluctuation, we regain the formalism of classical mechanics, i.e., $\tilde{p}(q;\xi)\rightarrow\tilde{p}_{\rm C}(q)=\partial_qS_{\rm C}$, so that the epistemic restriction also disappears: $\rho_{\tilde{p}}(q)\rightarrow\rho(q)$. This means that the global random variable $\xi$ provides the strength of the epistemic restriction, and requires that the fluctuation of $\xi$ should be microscopically small, so that it is practically ignorable in macroscopic regime. In this limit, $q$ sampled from $\rho(q)$ practically no longer carries any information about the conjugate momentum field $\tilde{p}_{\rm C}(q)$, and the Cram\'er-Rao inequality is no longer relevant. 

\section{Quantum calculus from best estimation of momentum given information on positions under epistemic restriction}

\subsection{Estimation error and information trade-off \label{Best estimation, estimation error, and information trade-off}}

To proceed, we need to know, given $q$, the error of each single shot estimation of $\tilde{p}(q;\xi)$ with the weakly unbiased estimator $\partial_qS(q)$, i.e. $\epsilon_p(q;\xi)\doteq \tilde{p}(q;\xi)-\partial_qS(q)$. This estimation error must reflect the actual microscopic physics underlying the momentum field $\tilde{p}(q;\xi)$ to be estimated, hence, the choice must be confronted with empirical evidences. It is first natural to demand that the estimation error is vanishing in the formal limit $\xi\rightarrow 0$, so that we regain classical mechanics. It must also reasonably satisfy the weak unbiased condition implying the Cram\'er-Rao inequality of Eq. (\ref{Cramer-Rao bound for momentum estimation}). Next, given $\xi$, it is intuitive to assume that the errors of estimating momentum fields arising in independent preparations must also be independent. Finally, we require that it leads to the best possible estimation, in the sense that the estimator $\partial_qS(q)$ must minimize the associated MS error measuring the accuracy of the estimation. 

To this end, we postulate that the error in a single shot estimation takes the following form:
\begin{eqnarray}
\epsilon_p(q;\xi)=\tilde{p}(q;\xi)-\partial_qS(q)=\frac{\xi}{2}\partial_q\ln\rho_{\tilde{p}}(q).
\label{estimation error}
\end{eqnarray}
As required, it is vanishing in the limit $\xi\rightarrow 0$. It is also clearly weakly unbiased, i.e., for any $\xi$, the error is indeed on average vanishing: $\int{\rm d}q~\tilde{p}(q;\xi)\rho_{\tilde{p}}(q)-\int{\rm d}q~\partial_qS(q)\rho_{\tilde{p}}(q)=\frac{\xi}{2}\int{\rm d}q~\partial_q\rho_{\tilde{p}}(q)=0$, where we have assumed that $\rho_{\tilde{p}}(q)$ is vanishing at infinity. Let us assume that $\xi$ fluctuates randomly on a microscopic time scale with a probability density $\chi(\xi)$ such that its first and second moments are constant in $(q,t)$, given respectively by \cite{Agung-Daniel model}:
\begin{eqnarray}
\overline{\xi}\doteq\int{\rm d}\xi~\xi~\chi(\xi)=0\hspace{2mm}\&\hspace{2mm}\overline{\xi^2}=\hbar^2. 
\label{Planck constant}
\end{eqnarray} 
The left equation in (\ref{Planck constant}) ensures that the estimation satisfies a reasonable independence condition that for any $q$, the estimation error $\epsilon_p(q;\xi)=\frac{\xi}{2}\partial_q\ln\rho_{\tilde{p}}(q)$ is uncorrelated with the estimator $\partial_qS(q)$, i.e., one has $\overline{(\partial_qS(q))(\epsilon_p(q;\xi))}=\frac{1}{2}\overline{\xi}\partial_qS \partial_q\ln\rho_{\tilde{p}}=0$. On the other hand, the right equation in (\ref{Planck constant}) says that the strength of the estimation error is on the order of Planck constant, so that for macroscopic systems it is practically ignorable, as required. In Subsection \ref{Quantum superposition and non-factorizable wave function} we show that, given $\xi$, it indeed leads to independent estimation errors for independent preparations. That the above choice of estimation error, i.e., Eqs. (\ref{estimation error}) and (\ref{Planck constant}), tally with empirical evidences and provide a scheme for best estimation minimizing the MS error, will be clarified in the next Subsection. 

As one crucial insight for the above choice, we note that Eqs. (\ref{estimation error}) and (\ref{Planck constant}) implies an information trade-off between the agent's knowledge about momentum and position. To see this, computing the MS error of the estimation of momentum field arising in a preparation one obtains:  
\begin{eqnarray}
\mathcal{E}_p^2&\doteq&\int{\rm d}q{\rm d}\xi~\epsilon_p(q;\xi)^2\chi(\xi)\rho_{\tilde{p}}(q)\nonumber\\
&=&\frac{\hbar^2}{4}\int{\rm d}q\big(\partial_q\ln\rho_{\tilde{p}}\big)^2\rho_{\tilde{p}}(q)=\frac{\hbar^2}{4}J_q, 
\label{the seed of quantum uncertainty}
\end{eqnarray}
where $J_q$ is the Fisher information about the mean position $q_o\doteq\int{\rm d}q q\rho_{\tilde{p}}(q)$ contained in $\rho_{\tilde{p}}(q)$, defined as \cite{Papoulis and Pillai book on probability and statistics,Cover-Thomas book} 
\begin{eqnarray}
J_q\doteq\int{\rm d}q\big(\partial_q\ln\rho_{\tilde{p}}(q)\big)^2\rho_{\tilde{p}}(q). 
\label{Fisher information of position}
\end{eqnarray}
One can see that Eq. (\ref{the seed of quantum uncertainty}) describes a trade-off | on the order of Planck constant | between the agent's estimation about the momentum and her/his information about the conjugate position, in an exact form. Namely, the smaller (larger) the MS error $\mathcal{E}_p^2$, i.e., the sharper (poorer) the agent's estimation about the momentum field, the smaller (larger) the Fisher information $J_q$, i.e., the poorer (better) her/his knowledge about the mean position. In this sense, one might say that Planck constant $\hbar$ makes the agent's information about momentum inseparable from her/his information about position, in the same spirit that the constant speed of light $c$ for all observers makes space and time inseparable in relativity theory. The above information trade-off clearly already reveals the essence of Heisenberg uncertainty principle from (classical) information theoretical view point. We shall show in Subsection \ref{Informational interpretation of quantum uncertainty, coherent superposition} that Eq. (\ref{the seed of quantum uncertainty}) indeed entails quantum uncertainty relation. 

\subsection{Emergent quantum calculus \label{Emergent quantum calculus}}

To show that the above choice of estimation error complies with empirical evidences in microscopic worlds, we note that Eqs. (\ref{estimation error}) and (\ref{Planck constant}) comprise the specific epistemic restriction we postulated in Ref. \cite{Agung-Daniel model} to reconstruct the mathematical formalism of spin-less non-relativistic quantum mechanics. For later convenient, we rewrite Eq. (\ref{estimation error}) as 
\begin{eqnarray}
\tilde{p}(q;\xi)=\partial_qS (q)+\frac{\xi}{2}\partial_q\ln\rho_{\tilde{p}}(q). 
\label{fundamental epistemic restriction}
\end{eqnarray}
First, the random momentum field of Eq. (\ref{fundamental epistemic restriction}) implies the following `epistemically restricted' phase space distribution \cite{Agung-Daniel model,Agung ERPS representation}:
\begin{eqnarray}
{\mathbb P}_{\{S ,\rho_{\tilde{p}}\}}(p,q|\xi)&=&\delta\big(p-\tilde{p}(q;\xi)\big)\rho_{\tilde{p}}(q)\nonumber\\
&=&\delta\Big(p-\partial_qS(q)-\frac{\xi}{2}\partial_q\ln\rho_{\tilde{p}}(q)\Big)\rho_{\tilde{p}}(q). 
\label{epistemically restricted phase space distribution}
\end{eqnarray}
It was then shown in Ref. \cite{Agung-Daniel model} that the ensemble average of any classical physical quantity $O(q,p)$ up to second order in $p$, over the phase space distribution of Eqs. (\ref{epistemically restricted phase space distribution}) and (\ref{Planck constant}), is equal to the quantum mechanics expectation value as
\begin{eqnarray}
\braket{O}_{\{S ,\rho_{\tilde{p}}\}}&\doteq&\int{\rm d}q{\rm d}\xi{\rm d}p~O(q,p){\mathbb P}_{\{S ,\rho_{\tilde{p}}\}}(p,q|\xi)\chi(\xi)\nonumber\\
&=&\braket{\psi|\hat{O}|\psi}.
\label{optical equivalence}
\end{eqnarray} 
Here, $\hat{O}$ is a Hermitian operator taking the same form as that obtained by applying the standard Dirac canonical quantization scheme to $O(q,p)$ with a specific ordering of operators, and the wave function $\psi(q)=\braket{q|\psi}$ is defined as
\begin{eqnarray}
\psi(q,t)\doteq\sqrt{\rho_{\tilde{p}}(q,t)}\exp\big(iS (q,t)/\hbar\big).
\label{wave function}
\end{eqnarray}
See Section Methods in Ref. \cite{Agung-Daniel model} for a proof. Equation (\ref{wave function}) implies that Born's statistical interpretation of wave function is valid by construction: 
\begin{eqnarray}
\rho_{\tilde{p}}(q,t)=|\psi(q,t)|^2.
\label{Born's rule}
\end{eqnarray} 

Further, imposing conservation of trajectories (or conservation of the probability current) manifested by the continuity equation, i.e.: $\partial_t\rho_{\tilde{p}}+\partial_q(\dot{q}\rho_{\tilde{p}})=0$, where $\dot{q}\doteq{\rm d}q/{\rm d}t$, and conservation of average energy, i.e.: $({\rm d}/{\rm d}t)\braket{H}_{\{S ,\rho_{\tilde{p}}\}}=0$, one can show that the time evolution of the wave function defined in Eq. (\ref{wave function}) must follow the unitary Schr\"odinger equation \cite{Agung-Daniel model}:   
\begin{eqnarray}
i\hbar\frac{{\rm d}}{{\rm d}t}\ket{\psi}=\hat{H}\ket{\psi}. 
\label{Schroedinger equation}
\end{eqnarray}
Here, $\hat{H}$ is the quantum Hamiltonian, again having the same form as that obtained by applying the standard canonical quantization to the classical Hamiltonian $H(q,p)$ with a specific ordering of operators. 

The above results can be extended straightforwardly to systems with many degrees of freedom, including interacting subsystems, generating quantum entanglement. In this case, as discussed in Ref. \cite{Agung-Daniel model}, to obtain Eqs. (\ref{optical equivalence}) and (\ref{Schroedinger equation}) for  Hamiltonian with cross terms of momentum between different degrees of freedom (as, e.g., arising in the measurement of momentum discussed in Appendix \ref{An illustration of wave function collapse in measurement momentum as Bayesian updating}), $\xi$ must indeed be global-nonseparable. Otherwise, if $\xi$ is separable, instead of quantum expectation value of energy and the Schr\"odinger equation, we regain respectively the conventional classical statistical average of energy and the classical Hamilton-Jacobi equation. Finally, applying the Schr\"odinger equation to a measurement interaction, and combining with Eq. (\ref{Born's rule}), noting that the system and measurement device in the model always follow a definite trajectory, one can derive the Born's rule for the statistics of measurement outcomes \cite{Agung-Daniel model}.  

Let us prove that the choice of estimation error of Eqs. (\ref{estimation error}) and (\ref{Planck constant}) implies that the weakly unbiased estimator $\partial_qS(q)$ best estimates $\tilde{p}(q;\xi)$ as claimed in the previous section. To this end, given information on $q$ obtained in measurement, we assume a general estimator $T_p(q)$ for $\tilde{p}(q;\xi)$, and compute the associated MS error to first obtain $\int{\rm d}q{\rm d}\xi\big(\tilde{p}(q;\xi)-T_p(q)\big)^2\chi(\xi)\rho_{\tilde{p}}(q)=\braket{\psi|(\hat{p}-T_p(q))^2|\psi}$, where we have used Eq. (\ref{optical equivalence}). The right hand side can be further expanded to give $\int{\rm d}q{\rm d}\xi\big(\tilde{p}(q;\xi)-T_p(q)\big)^2\chi(\xi)\rho_{\tilde{p}}(q)=\braket{\psi|\hat{p}^2|\psi}+\int{\rm d}q\rho_{\tilde{p}}(q)\big((T_p(q)-\partial_qS(q))^2-(\partial_qS(q))^2\big)$, where we have used Eq. (\ref{wave function}). One can then see that it is minimized when $T_p(q)=\partial_qS(q)$ as claimed. We note that such {\it best} estimation is not necessarily {\it efficient}, i.e., it does not necessarily saturate the Cram\'er-Rao inequality of Eq. (\ref{Cramer-Rao bound for momentum estimation}). We shall see in Subsection \ref{Informational interpretation of quantum uncertainty, coherent superposition} and Appendix \ref{Gaussian wave function describes an efficient estimation of momentum} that the Cram\'er-Rao bound of Eq. (\ref{Cramer-Rao bound for momentum estimation}) is attained by certain class of `Gaussian preparations'. Finally, it is interesting that the form of estimation error for {\it best} estimation of Eq. (\ref{estimation error}) has a similar form as that for {\it efficient} estimation of Eq. (\ref{Cramer-Rao optimal condition}), except that there the derivative is with respect to $\tilde{p}$ instead of $q$. 

\subsection{Wave function represents an agent's best estimation of momentum given information on positions \label{Wave function represents an agent's best estimation of momentum given information on the conjugate position}}

We first note that we still keep the conjecture put in Ref. \cite{Agung-Daniel model} that there is an ontic or real dynamics underlying the statistical model, i.e., the dynamics of the actual random momentum field for an individual system. This dynamics of the momentum field must {\it fundamentally} restrict the dynamics of the distribution of positions; this is the essence of the epistemic restriction. We still do not know the exact nature of the ontic dynamics of the momentum field for individual system, and why, unlike in conventional classical mechanics, such dynamics of momentum field fundamentally restricts (thus must be irreducibly correlated with) that of the distribution of positions. 

We then decompose the random momentum field into two terms as in Eq. (\ref{fundamental epistemic restriction}). We emphasize however that this decomposition is not physical (realist). In particular, it is not a decomposition of the momentum field $\tilde{p}(q;\xi)$ into a deterministic (classical) component $\partial_qS(q)$ subjected to a stochastic physical noise $\epsilon_p(q;\xi)=\frac{\xi}{2}\partial_{q}\ln\rho_{\tilde{p}}(q)$ as in Nelson stochastic mechanics \cite{Nelson stochastic mechanics}. Such a physical decomposition will have to face a conceptual difficulty that the momentum field $\tilde{p}(q;\xi)$ (an ontic variable) might be causally affected by the probability distribution of position $\rho_{\tilde{p}}(q)$ (an epistemic quantity). Rather, the decomposition in Eq. (\ref{fundamental epistemic restriction}) is epistemic or informational: namely, given information on $q$, an agent must see the term $\partial_qS(q)$ as her/his weakly unbiased best estimate of $\tilde{p}(q;\xi)$, with the random estimation error $\epsilon_p(q;\xi)=\frac{\xi}{2}\partial_{q}\ln\rho_{\tilde{p}}(q)$. Hence, the decomposition in Eq. (\ref{fundamental epistemic restriction}) does not happen in real physical space, but in the agent's mind. 

We have shown that the epistemic decomposition of momentum field of Eq. (\ref{fundamental epistemic restriction}) leads to the quantum kinematics of wave functions and Hermitian operators defined in Eqs. (\ref{optical equivalence}) and (\ref{wave function}) \cite{Agung-Daniel model}. In this sense, we can regard the epistemic decomposition  to provide a kind of kinematics restriction \cite{Bub pre-dynamics probabilistic constraint}, in an analogous way that the Minkowski space provides a spacetime constraint in relativity theory. We may therefore conclude within the above epistemic reconstruction based on estimation scheme, that the wave function $\psi(q)$ defined in Eq. (\ref{wave function}) implying Eqs. (\ref{optical equivalence}) and (\ref{Schroedinger equation}), does not represent the actual objective physical state | i.e., the random momentum field | arising in the preparation. Rather, $\psi(q)$ should be interpreted to summarize the agent's weakly unbiased best estimate of the momentum and the corresponding estimation error, based on information on the conjugate position, under epistemic restriction. Quantum wave function is therefore agent-dependent, i.e., relative to the agent making the estimation given prior information about her/his preparation. It lives in the agent's mind, rather than in physical space. And, since the wave function describes an estimation, it must therefore be meaningful even for an individual system. At this point, we note that Pusey, Barrett and Rudolph  (PBR) have recently devised a theorem which shows the conceptual difficulty to hold a realist interpretation of quantum mechanics with such an epistemic wave function \cite{PBR theorem}. We shall show at the end of Subsection \ref{Quantum superposition and non-factorizable wave function} that the statistical model with epistemic wave function proposed in this paper violates the assumption of preparation independence underlying the PBR theorem.  

Let us make more precise the above epistemic interpretation of wave function. Suppose an agent makes an experimental preparation by varying some controllable macroscopic parameters to select a specific wave function $\psi(q)$ among those allowed by the experimental arrangement. What does the selected $\psi(q)$ mean? Within the above epistemic reconstruction based on estimation scheme, noting Eqs. (\ref{fundamental epistemic restriction}) (or (\ref{estimation error})) and (\ref{wave function}), a preparation quantum mechanically represented by $\psi(q)$ means that the system undergoes a momentum fluctuation $\tilde{p}(q;\xi)$ such that given information on $q$, the agent's weakly unbiased best estimate of the momentum field at $q$, denoted by $\overline{p}(q)$, and the associated estimation error $\epsilon_p(q;\xi)=\tilde{p}(q;\xi)-\overline{p}(q)$, are determined by $\psi(q)$, respectively, as 
\begin{eqnarray}
\overline{p}(q)=\partial_qS(q)=\frac{\hbar}{2i}\big(\partial_q\ln\psi(q)-\partial_q\ln\psi^*(q)\big),\nonumber\\
\epsilon_p(q;\xi)=\frac{\xi}{2}\partial_q\ln\rho_{\tilde{p}}(q)=\frac{\xi}{2}\partial_q\ln\big|\psi(q)\big|^2.~~~
\label{efficient estimation of momentum and random error} 
\end{eqnarray}    

We show that the above epistemic interpretation allows for a conceptually transparent quantum-classical boundary and transition. First, in the physical regime wherein the magnitude of estimation error is much smaller than the magnitude of the estimator, i.e., $|\overline{p}|=|\partial_qS|\gg |\epsilon_p|=|\frac{\xi}{2}\partial_q\ln\rho_{\tilde{p}}|$, the second term on the right hand side of Eq. (\ref{fundamental epistemic restriction}) can be practically ignored and we regain classical mechanics relation $\tilde{p}\approx\overline{p}=\partial_qS$ of Eq. (\ref{Hamilton-Jacobi condition}) (by identifying $S$ with $S_{\rm C}$ and $\tilde{p}$ with $\tilde{p}_{\rm C}$). This is approximately true in the macroscopic regime due to Eq. (\ref{Planck constant}) that the strength of the estimation error is on the order of $\hbar$. Hence, in the macroscopic regime, given $q$, an agent can in principle prepare the momentum $p=\tilde{p}(q)\approx\partial_qS(q)$ of the system sharply with practically negligible error.     

\subsection{Quantum canonical commutation relation from estimation of momentum given the positions under epistemic restriction \label{Estimation of momentum field under epistemic restriction and quantum incompatibility}}

One of the basic mathematical axioms of quantum calculus is the canonical commutation relation between position and momentum operators, i.e., $[\hat{q},\hat{p}]\doteq\hat{q}\hat{p}-\hat{p}\hat{q}=i\hbar$, sometimes regarded as the ``fundamental quantum condition'' \cite{Dirac book}. This commutation relation underlies the noncommutative structure of quantum observables, and is at the root of many important mathematical results of quantum mechanics, including the violation of Bell's inequality \cite{Bell's theorem} as suggested in Refs. \cite{Fine-Bell theorem,Cavalcanti non-commutativity and Bell inequality,Griffiths non-commutativity and Bell inequality}. It is therefore instructive to ask how the canonical commutation relation is reflected within the epistemic reconstruction of quantum mechanics based on the scheme of estimation under epistemic restriction discussed above. 

To study this question, we express the MS error of the estimation of momentum field given in Eq. (\ref{the seed of quantum uncertainty}) using the language of quantum calculus in complex Hilbert space as: 
\begin{eqnarray}
\mathcal{E}_p^2&=&\frac{\hbar^2}{4}\int{\rm d}q\big(\partial_q\ln\rho_{\tilde{p}}\big)^2\rho_{\tilde{p}}(q)\nonumber\\
&=&\int {\rm d}q\Big(\frac{\braket{\psi|[\hat{\pi}_q,\hat{p}]|\psi}}{2i|\braket{q|\psi}|^2}\Big)^2|\braket{q|\psi}|^2,
\label{MS momentum vs canonical commutation relation}
\end{eqnarray}
where $\hat{\pi}_q\doteq |q\rangle\langle q|$ is the projector onto $\ket{q}$. One can then see that if, unlike our statistical model, $\mathcal{E}_p^2$ is vanishing for all preparations, then we must have $[\hat{\pi}_q,\hat{p}]=0$, in contradiction with canonical commutation relation. Further, let us estimate the mean position $q_o$ with the unbiased estimator $q$, so that the MS error reads $\mathcal{E}_q^2\doteq\int{\rm d}q(q-q_o)^2\rho_{\tilde{p}}(q)=\int{\rm d}q(q-q_o)^2|\braket{q|\psi}|^2$. Multiplying this to Eq. (\ref{MS momentum vs canonical commutation relation}), and applying the Cauchy-Schwarz inequality, one obtains, after a straightforward manipulation, a Robertson-like uncertainty relation 
\begin{eqnarray}
\mathcal{E}_p^2\mathcal{E}_q^2\ge\frac{1}{4}\big|\braket{\psi|[\hat{q},\hat{p}]|\psi}\big|^2, 
\label{canonical commutation relation from estimations}
\end{eqnarray}
where $\hat{q}\doteq\int{\rm d}qq|q\rangle\langle q|$. 

In this sense, we may therefore conclude that the epistemic decomposition of the momentum field into the weakly unbiased best estimate and its error of Eq. (\ref{fundamental epistemic restriction}), which together with Eq. (\ref{Planck constant}) implies the information trade-off in Eq. (\ref{the seed of quantum uncertainty}), gives the informational origin of the quantum canonical commutation relation in complex Hilbert space formalism of quantum mechanics. Indeed, we shall show in Subsection \ref{Informational interpretation of quantum uncertainty, coherent superposition} without using quantum calculus that the right hand side of Eq. (\ref{canonical commutation relation from estimations}) is given by $\hbar^2/4$. 

\subsection{Estimation of momentum based on information on position versus momentum weak value at a given position \label{Best estimation under epistemic restriction and quantum weak values}}

We show that the above scheme of estimation is deeply related to the intriguing concept of weak value obtained from weak measurement over pre- and post-selected ensemble \cite{Aharonov weak value}. Consider an ensemble of repeated weak measurement of momentum in such a way that it only gently perturbs the initial preparation, and followed by a post-selection on (a sub-ensemble compatible with) the conjugate position $q$. It can be shown within the operational formalism of quantum mechanics  \cite{Lundeen complex weak value,Jozsa complex weak value} that the average shift of the position of the pointer of the measuring device is proportional to the real part of the weak momentum value at $q$ given by: ${\rm Re}\{\frac{\braket{q|\hat{p}|\psi}}{\braket{q|\psi}}\}=\partial_qS(q)$, where we have used Eq. (\ref{wave function}). Moreover, the average shift of the pointer momentum is proportional to the imaginary part of the weak momentum value at $q$: ${\rm Im}\{\frac{\braket{q|\hat{p}|\psi}}{\braket{q|\psi}}\}=-\frac{\hbar}{2}\partial_q\ln\rho_{\tilde{p}}(q)$. Noting this, first, the epistemically restricted phase space distribution of Eq. (\ref{epistemically restricted phase space distribution}) associated with $\psi(q)=\sqrt{\rho_{\tilde{p}}(q)}\exp(iS(q)/\hbar)$ is not just a mathematical artefact, but can be determined experimentally via weak momentum value as ${\mathbb P}_{\{S ,\rho_{\tilde{p}}\}}(p,q|\xi)=\delta\big(p-{\rm Re}\{\frac{\braket{q|\hat{p}|\psi}}{\braket{q|\psi}}\}+\frac{\xi}{\hbar}{\rm Im}\{\frac{\braket{q|\hat{p}|\psi}}{\braket{q|\psi}}\}\big)|\braket{q|\psi}|^2$ \cite{Agung ERPS representation}. Moreover, the weakly unbiased best estimate of the momentum based on information of position of Eq. (\ref{efficient estimation of momentum and random error}) and the associated MS error of Eq. (\ref{the seed of quantum uncertainty}), can thus be experimentally probed via weak measurement of momentum followed by post-selection on position, respectively as 
\begin{eqnarray}
\overline{p}(q)&=&\partial_qS={\rm Re}\Big\{\frac{\braket{q|\hat{p}|\psi}}{\braket{q|\psi}}\Big\},\nonumber\\
\mathcal{E}_p^2&=&\int{\rm d}q\Big({\rm Im}\Big\{\frac{\braket{q|\hat{p}|\psi}}{\braket{q|\psi}}\Big\}\Big)^2|\braket{q|\psi}|^2.
\label{probing efficient estimation and the MS error via weak value}
\end{eqnarray}
This shows that the outcome of weak momentum measurement with position post-selection also represents the agent's weakly unbiased best estimate of momentum given position, rather than revealing what exist objectively prior to observation. 

Let us mention that in Ref. \cite{Hall exact uncertainty and best estimation}, working within the standard formalism of quantum mechanics, Hall argued that, for a given pure quantum state written in polar form as in Eq. (\ref{wave function}), an operator defined as $\hat{p}_{\rm C}\doteq\int{\rm d}q~\partial_qS(q)\ket{q}\bra{q}$ can be seen as the `best classical estimate' of $\hat{p}$ compatible with $\hat{q}$, minimizing the quantum MS error defined as the quantum statistical deviation between $\hat{p}$ and $\hat{p}_{\rm C}$ over $\ket{\psi}$. In this way, he derived from quantum mechanics a similar result as in Eq. (\ref{the seed of quantum uncertainty}). Namely, he defined $\delta X\doteq J_q^{-1/2}$ as `Fisher length', and $\Delta P_{nc}\doteq\sqrt{\braket{\psi|(\hat{p}-\hat{p}_{\rm C})^2|\psi}}=\sqrt{\mathcal{E}_p^2}$ as the strength of `nonclassical momentum fluctuation', and write Eq. (\ref{the seed of quantum uncertainty}) as $\Delta P_{nc}\delta X=\hbar/2$, calling it an `exact uncertainty relation'. Johansen in Ref. \cite{Johansen weak value as Bayesian efficient estimator} further showed that $\hat{p}_{\rm C}$ defined above can be interpreted as the best estimate of $\hat{p}$ over a pre-selected ensemble represented by a quantum state $\ket{\psi}$ and followed by post-selection of a sub-ensemble compatible with $q$ within weak measurement scheme, and derived Eq. (\ref{probing efficient estimation and the MS error via weak value}). In Ref. \cite{Hall best estimate} Hall later argued that $\partial_qS(q)$ should be seen as the best estimate of $\hat{p}$ based on measurement of $\hat{q}$.  

Our interpretation of wave function in terms of estimation of momentum given information on position thus somehow combines all the results reported by the above authors, while providing a new insight based on the notion of epistemic restriction proposed in Ref. \cite{Agung-Daniel model}. We note first that these authors worked within the standard formalism of quantum mechanics rather than reconstructing it from scratch. By contrast, here we start from a weakly unbiased, best estimation of momentum given information on position, in the presence of epistemic restriction, to reconstruct quantum mechanics. Further, in \cite{Hall exact uncertainty and best estimation,Johansen weak value as Bayesian efficient estimator,Hall best estimate}, while the authors gave the form of the quantum mechanical MS error in terms of quantum statistical deviation, they did not offer the form of the error arising at each single shot of estimation. By contrast, we have provided the exact functional form of the random estimation error $\epsilon_p(q;\xi)$ for each single shot of estimation given in Eq. (\ref{efficient estimation of momentum and random error}) (or Eq. (\ref{estimation error})), introducing a global-nonseparable random variable $\xi$. We emphasize that it is the introduction of $\xi$ satisfying Eq. (\ref{Planck constant}) parameterizing the error function $\epsilon_p(q;\xi)$ in Eq. (\ref{efficient estimation of momentum and random error}) which leads to the derivation of abstract mathematical rules of quantum mechanics given in Eqs. (\ref{optical equivalence}) - (\ref{Schroedinger equation}). 

Next, in \cite{Wiseman Bohmian mechanics} Wiseman introduced the notion of  `naively observable' average velocity at $q$, denoted as $v_{\rm W}(q)$, and interpreted it operationally using quantum weak value. He showed that, for a particle of mass $m$ in a scalar potential, the naively observable average velocity determined operationally as weak velocity value, is exactly equal to the Bohmian velocity \cite{Bohm pilot-wave theory}, i.e., $v_{\rm W}(q)={\rm Re}\{\frac{\braket{q|\hat{p}|\psi}}{\braket{q|\psi}}\}/m=\partial_qS(q)/m$. This thus operationally justifies the unique choice of the Bohmian velocity in terms of weak velocity value. Noting Eq. (\ref{efficient estimation of momentum and random error}), one can however see within our epistemic reconstruction that Wiseman naively observable average velocity is also equal to the weakly unbiased best estimate of velocity given information on $q$, i. e., $\overline{v}(q)\doteq\overline{p}(q)/m=\partial_qS(q)/m=v_{\rm W}(q)$. This observation suggests that Wiseman naively observable average velocity and Bohmian velocity should be read subjectively as weakly unbiased best estimate of velocity given information on $q$, rather than revealing an agent-independent objective velocity of the particle as favored by Bohmian mechanics.

\section{Epistemic interpretation of quantum mechanics \label{Informational interpretation of quantum uncertainty, coherent superposition}}

\subsection{Informational origin of quantum uncertainty and complementarity \label{Informational interpretation of quantum uncertainty, coherent superposition}}

Consider an agent preparing a system quantum mechanically represented by a general wave function $\psi(q)$. As discussed in the previous section, it means that the system experiences a random momentum fluctuation $\tilde{p}(q;\xi)$ fundamentally restricting the allowed distribution of positions $\rho_{\tilde{p}}(q)$, such that the MS error of the agent's weakly unbiased best estimate of the momentum, given the position, with the estimator $\overline{p}(q)=\partial_qS(q)$, must satisfy Eq. (\ref{the seed of quantum uncertainty}). On the other hand, in a general unbiased estimation of mean position $q_o$, the corresponding MS error must satisfy the Cram\'er-Rao inequality associated with the estimation \cite{Papoulis and Pillai book on probability and statistics,Cover-Thomas book}. Taking $q$ as the unbiased estimator for $q_o$ one therefore has 
\begin{eqnarray}
\mathcal{E}_q^2\doteq\int{\rm d}q(q-q_o)^2\rho_{\tilde{p}}(q)\ge \frac{1}{J_q}, 
\label{Cramer-Rao inequality for position}
\end{eqnarray}
where $J_q$ is the Fisher information about $q_o$ defined in Eq. (\ref{Fisher information of position}). Multiplying both sides with $\mathcal{E}_p^2$ and using Eq. (\ref{the seed of quantum uncertainty}), one finally obtains, 
\begin{eqnarray}
\mathcal{E}_p^2\mathcal{E}_q^2\ge \frac{\hbar^2}{4},
\label{trade off between MS error of position and momentum}
\end{eqnarray}
describing a trade-off between the MS errors of the joint/simultaneous weakly unbiased best estimation of momentum given $q$ with the estimator $\partial_qS(q)$, and an unbiased estimation of mean position $q_o$ with the estimator $q$. Note that Eq. (\ref{trade off between MS error of position and momentum}) is linked to Eq. (\ref{canonical commutation relation from estimations}) via the canonical commutation relation. Further, as shown in Appendix \ref{Efficient estimation of mean position and Gaussian distribution of error}, the estimation of mean position $q_o$ with the estimator $q$ is efficient saturating the Cram\'er-Rao inequality of Eq. (\ref{Cramer-Rao inequality for position}), when the probability distribution of estimation error $q-q_o$ takes the form of a Gaussian $\rho_{\tilde{p}}(q)=\frac{1}{\sqrt{2\pi\sigma_q^2}}e^{-\frac{(q-q_o)^2}{2\sigma_q^2}}$. For Gaussian distribution of $q-q_o$, the inequality in Eq. (\ref{trade off between MS error of position and momentum}) is therefore also saturated: $\mathcal{E}_p^2\mathcal{E}_q^2=\hbar^2/4$. 

Hence, no preparation gives the agent sharp weakly unbiased best estimate of the momentum field and sharp unbiased estimate of mean position, simultaneously, based on information on $q$: the product of their MS errors is bounded from below by the Planck constant, the strength of the fluctuation of $\xi$. It thus somewhat reflects Bohr's principle of complementarity: Bohr's fundamental limitations in {\it jointly defining} complementary quantities (like position and momentum) within the context of a measurement \cite{Bohr Q epistemic}, is reflected within our epistemic interpretation by the agent's limitation in {\it jointly estimating} complementarity quantities resulting in a preparation. 

To connect the MS errors trade-off of Eq. (\ref{trade off between MS error of position and momentum}) with the quantum uncertainty relation, we need to interpret the quantum variance of position and momentum operators within the above interpretative framework based on estimation scheme. First, using Eq. (\ref{wave function}), we find that the quantum variance $\sigma_{\hat q}^2$ of position operator $\hat{q}$ is mathematically equal to the MS error of estimation of mean position $\mathcal{E}_q^2$ defined in Eq. (\ref{Cramer-Rao inequality for position}), i.e.: $\sigma_{\hat q}^2=\braket{\psi|(\hat{q}-\braket{\psi|\hat{q}|\psi})^2|\psi}=\int{\rm d}q(q-q_o)^2\rho_{\tilde{p}}(q)=\mathcal{E}_q^2$. Moreover, we show in Appendix \ref{Decomposition of quantum variance of momentum into the accuracy and impression of best estimation of momentum} that the quantum variance $\sigma_{\hat p}^2$ of the momentum operator $\hat{p}$ can be decomposed as: $\sigma_{\hat p}^2=\braket{\psi|(\hat{p}-\braket{\psi|\hat{p}|\psi})^2|\psi}=\mathcal{E}_p^2+\Delta_p^2$, where $\Delta_p^2\doteq\int{\rm d}q\big(\partial_qS(q)-\int{\rm d}q'\partial_{q'}S(q')\rho_{\tilde{p}}(q')\big)^2\rho_{\tilde{p}}(q)$ is the dispersion of the estimator $\partial_qS(q)$. Hence, within the estimation scheme, $\sigma_{\hat p}^2$ should be interpreted as the sum of the MS error of estimation of momentum (i.e., inaccuracy of estimation) and the dispersion of the estimator (i.e., imprecision of estimation) \cite{Hall best estimate}. Combining these two equations, one finally obtains, by the virtue of Eq. (\ref{trade off between MS error of position and momentum}), and noting the fact that $\Delta_p^2\ge 0$, the Heinseberg-Kennard uncertainty relation
\begin{eqnarray}
\sigma_{\hat p}^2\sigma_{\hat q}^2&=&\mathcal{E}_p^2\mathcal{E}_q^2+\Delta_p^2\mathcal{E}_q^2\ge\frac{\hbar^2}{4} +\Delta_p^2\mathcal{E}_q^2\ge\frac{\hbar^2}{4}. 
\label{Heisenberg-Kennard uncertainty relation}
\end{eqnarray}
Note that the Heisenberg-Kennard uncertainty relation for preparation underlies the uncertainty relations arising in measurement, either describing the trade-off between error-disturbance \cite{Ozawa error-disturbance UR}, or between errors in joint measurement \cite{Ozawa joint-measurement UR,Hall best estimate}.  

We have shown above that the first inequality in Eq. (\ref{Heisenberg-Kennard uncertainty relation}) is saturated when the estimation of $q_o$ by $q$ is efficient, so that the distribution of $q-q_o$ takes a Gaussian. On the other hand, one can see from Eq. (\ref{Heisenberg-Kennard uncertainty relation}) that the second inequality is saturated when $\Delta_p^2=0$. This can be solved to give $S(q)=p_oq$, where $p_o$ is an arbitrary real constant. Combining these two facts, and noting Eq. (\ref{wave function}), both the inequalities in Eq. (\ref{Heisenberg-Kennard uncertainty relation}) are therefore saturated when the wave function associated with the preparation is given by the Gaussian wave function: 
\begin{eqnarray}
\psi(q)=\Big(\frac{1}{2\pi\sigma_q^2}\Big)^{1/4}e^{-\frac{(q-q_o)^2}{4\sigma_q^2}+ip_oq/\hbar}.
\label{Gaussian wave function}
\end{eqnarray} 
We further show in Appendix \ref{Gaussian wave function describes an efficient estimation of momentum} that for this Gaussian preparation, the {\it best} estimate of the momentum field $\tilde{p}$ by the estimator $\overline{p}=\partial_qS(q)=p_o$ is also {\it efficient} saturating the Cram\'er-Rao inequality of Eq. (\ref{Cramer-Rao bound for momentum estimation}). In this respect, Gaussian wave function is thus special, that is, it describes a preparation wherein the agent's simultaneous estimations of the momentum field and mean position parameterizing $\rho_{\tilde{p}}(q)$ are both efficient, achieving the Cram\'er-Rao bounds of the associated MS errors.  These results show that quantum uncertainty relation is fundamentally related to the Cram\'er-Rao inequality limiting the agent's estimation.  

\subsection{Quantum superposition and non-factorizable wave function\label{Quantum superposition and non-factorizable wave function}}

Let us proceed to discuss a specific combination of preparations and its implication to the agent's estimation. Consider the superposition of two wave functions $\psi(q)\sim\psi_1(q)+\psi_2(q)$, where $\psi_j(q)=\sqrt{\rho_{\tilde{p}_j}(q)}e^{\frac{i}{\hbar}S_j(q)}$, $j=1,2$. First, assume that the two wave functions $\psi_j(q)$, $j=1,2$, do not overlap, i.e., $\Lambda_1\cap\Lambda_2=\emptyset$, where $\Lambda_j$ is the support of $\psi_j(q)$, $j=1,2$. In this case, information on $q$ is sufficient to discriminate unambigously the two wave functions $\psi_j(q)$, $j=1,2$, each represents the agent's best estimation of the momentum field arising in a distinct preparation. Moreover, as shown in Appendix \ref{Efficient estimation of momentum and its error for the superposition of two wave functions}, given $q\in\Lambda_j$, $j=1,2$, the agent's best estimate of the momentum field $\tilde{p}(q;\xi)$ associated with $\psi(q)\sim\psi_1(q)+\psi_2(q)$ is given by $\overline{p}(q)=\overline{p}_j(q)$ with the corresponding estimation error $\epsilon_p(q;\xi)=\epsilon_{p_j}(q;\xi)$. Here, $\overline{p}_j(q)\doteq\partial_qS_j(q)$ and $\epsilon_{p_j}(q;\xi)\doteq\frac{\xi}{2}\partial_q\ln\rho_{\tilde{p}_j}(q)$ are respectively the agent's best estimate of the momentum field $\tilde{p}_j(q)$ and the corresponding estimation error, associated with $\psi_j(q)$, $j=1,2$. For $q\in\Lambda_j$, $j=1,2$, we thus have $\tilde{p}(q;\xi)=\overline{p}(q)+\epsilon_p(q;\xi)=\overline{p}_j(q)+\epsilon_{p_j}(q;\xi)=\tilde{p}_j(q;\xi)$. Hence, any trajectory belonging to the momentum field $\tilde{p}$ associated with $\psi\sim\psi_1+\psi_2$, must also belong to either $\tilde{p}_1$ associated with $\psi_1$, or $\tilde{p}_2$ associated with $\psi_2$. In this sense, we say that $\tilde{p}$ associated with $\psi\sim\psi_1+\psi_2$ is compatible with $\tilde{p}_j$ associated with $\psi_j$, $j=1,2$. The ensemble of trajectories following $\tilde{p}$ associated with $\psi\sim\psi_1+\psi_2$ is thus a statistical mixture of those following $\tilde{p}_j$ associated with $\psi_j$, $j=1,2$. Accordingly, the agent should expect that the probability distribution of position is additively decomposable: $\rho_{\tilde{p}}(q)=\rho_{\tilde{p}_1}(q)+\rho_{\tilde{p}_2}(q)$, with no interference term, as predicted by quantum mechanics. The above observation shows that, if $\psi_1$ and $\psi_2$ are nonoverlapping, given $q$, the agent's weakly unbiased best estimation of momentum field represented by $\psi_j$, $j=1,2$, is sufficient to fully account for that represented by their superposition $\psi\sim\psi_1+\psi_2$. In this case, as intuitively expected, the MS error is additive: $\mathcal{E}_p^2[\psi]=\mathcal{E}_p^2[\psi_1]+\mathcal{E}_p^2[\psi_2]$. 

If instead $\psi_1(q)$ and $\psi_2(q)$ are overlapping on a non-empty set $\Lambda_{12}\doteq\Lambda_1\cap\Lambda_2\neq\emptyset$, then for each $q\in\Lambda_{12}$, the above conclusions do not apply. First, information on $q\in\Lambda_{12}$ is no longer sufficient for the agent to unambiguously distinguish the two preparations represented by $\psi_j$, $j=1,2$. What is the form of the momentum field $\tilde{p}(q;\xi)$ in the overlapping region? We note that this momentum field must respect the epistemic restriction of Eq. (\ref{fundamental epistemic restriction}). Inserting $\psi=\psi_1+\psi_2$ into Eqs. (\ref{fundamental epistemic restriction}) or (\ref{efficient estimation of momentum and random error}), one can show that at $q\in\Lambda_{12}$, the associated momentum field $\tilde{p}$ is not equal to either $\tilde{p}_1$ or $\tilde{p}_2$ (see Eq. (\ref{general formula for the efficient estimation and errors for superposition}) for the exact forms of $\overline{p}$, $\epsilon_p$, and thus $\tilde{p}=\overline{p}+\epsilon_p$ at $q\in\Lambda_{12}$). Namely, there are trajectories passing $q\in\Lambda_{12}$ belonging to the momentum field $\tilde{p}_j$, associated with $\psi_j$, $j=1,2$, which do not belong to the momentum field $\tilde{p}$ associated with $\psi\sim\psi_1+\psi_2$, and vice versa. Hence, they are no longer compatible: their sample spaces are incompatible. The ensemble of trajectories following $\tilde{p}$ associated with $\psi\sim\psi_1+\psi_2$ thus cannot in general be decomposed into a statistical mixture of ensembles of trajectories following $\tilde{p}_j$ associated with $\psi_j$, $j=1,2$. Accordingly, at $q\in\Lambda_{12}$, we must have $\rho_{\tilde{p}}(q)\neq\rho_{\tilde{p}_1}(q)+\rho_{\tilde{p}_2}(q)$, namely, there is an interference term as predicted by quantum mechanics. This result does not violate the usual law of total probability since the probabilities on the left and right hand sides are parameterized by different incompatible momentum fields (via epistemic restriction), i.e., the two sides refer to two incompatible contexts.  

Hence, interference term in position space arising in coherent superposition is implied by the presence of epistemic restriction (i.e., the restriction on the allowed probability distribution of position by the underlying momentum field), and the fact that in the overlapping region $\Lambda_{12}$, the momentum field $\tilde{p}$ associated with $\psi\sim\psi_1+\psi_2$ is not compatible with $\tilde{p}_j$ associated with $\psi_j$, $j=1,2$. This means that, at $q\in\Lambda_{12}$, the agent's best estimation of the momentum fields $\tilde{p}_1$ and $\tilde{p}_2$ respectively represented by $\psi_1$ and $\psi_2$, is not sufficient to fully account for her/his estimation of the momentum field $\tilde{p}$ represented by $\psi\sim\psi_1+\psi_2$. Accordingly, in this case, the MS error is not additively decomposable: $\mathcal{E}_p^2[\psi]\neq \mathcal{E}_p^2[\psi_1]+\mathcal{E}_p^2[\psi_2]$, as intuitively expected. A simple example is given by the superposition of two plane wave functions $\psi_1\sim e^{ip_oq/\hbar}$ and $\psi_2\sim e^{-ip_oq/\hbar}$, so that $\psi\sim\psi_1+\psi_2\sim\cos(p_oq/\hbar)$. While the agent's best estimation of momentum field arising in the two preparations, each represented by a plane wave, are sharp with vanishing MS errors, i.e., $\mathcal{E}_p^2[\psi_j]=0$, $j=1,2$, her/his best estimation of momentum field arising in the combination of the two preparations represented by the superposition of the two plane waves has a finite MS error i.e., $\mathcal{E}_p^2[\psi_1+\psi_2]=p_o^2>0$. This example also shows that quantum coherence in momentum basis reflects that the MS error of best estimation of momentum field is non-vanishing.  

A similar epistemic reading applies to separable and inseparable (entangled) wave functions of two or more subsystems. Consider first independent preparations of two subsystems, $A$ and $B$, with a spatial configuration $q=(q_A,q_B)$, so that quantum mechanically it is described by a separable total wave function $\psi(q_A,q_B)=\psi_A(q_A)\psi_B(q_B)$. In this case, noting Eq. (\ref{wave function}), the phase of the wave function is additively decomposable, $S (q_A,q_B)=S_A(q_A)+S_B(q_B)$, and the amplitude is separable, $\rho_{\tilde{p}}(q_A,q_B)=\rho_{\tilde{p}_A}(q_A)\rho_{\tilde{p}_B}(q_B)$, so that inserting into Eq. (\ref{efficient estimation of momentum and random error}), given the positions $q=(q_A,q_B)$, the agent's weakly unbiased best estimate of the momentum and its estimation error are separable, respectively given by: $\overline{p}_I(q_I)=\partial_{q_I}S_I(q_I)$ and $\epsilon_{p_I}(q_I;\xi)=\frac{\xi}{2}\partial_{q_I}\ln\rho_{\tilde{p}_I}(q_I)$, $I=A,B$. This shows that separable wave function represents {\it independent} pair of preparations so that information on $q_A(q_B)$ is sufficient for the agent to make {\it independent}, weakly unbiased, best estimation about the momentum field $\tilde{p}_A(\tilde{p}_B)$, i.e., the associated estimators and estimation errors are both {\it independent}. In this case, the MS error of joint estimation is additively decomposable: $\mathcal{E}_p^2[\psi]\doteq\sum_{I=A,B}\int{\rm d}q_A{\rm d}q_B{\rm d}\xi(\tilde{p}_I-\partial_{q_I}S)^2\chi(\xi)\rho_{\tilde{p}}(q)=\sum_{I=A,B}\int{\rm d}q_I{\rm d}\xi\big(\frac{\xi}{2}\partial_{q_I}\ln\rho_{\tilde{p}_I}(q_I)\big)^2\chi(\xi)\rho_{\tilde{p}_I}(q_I)=\mathcal{E}_p^2[\psi_A]+\mathcal{E}_p^2[\psi_B]$, as intuitively expected. On the other hand, when the wave function is entangled, i.e., $\psi(q_A,q_B)\neq\psi_A(q_A)\psi_B(q_B)$, the above conclusion is no longer valid. For example, information on $q_A$ is no longer sufficient for the agent to make weakly unbiased best estimation of the momentum field associated with subsystem $A$, i.e., $\overline{p}_A$ and/or $\epsilon_{p_A}$ now may depend on the position $q_B$ of subsystem $B$, even if the two subsystems are remotely separated from each other. Accordingly, the MS error is no longer additive. Entangled wave function thus represents correlated preparations due to interaction in the past, so that independent weakly unbiased best estimation is impossible. 

Next, within the model, the measurement outcome is determined by the initial position, initial momentum field, and importantly by a finite time fluctuation of $\xi(t)$, denoted as $[\xi(t)]$ (see Appendix \ref{An illustration of wave function collapse in measurement momentum as Bayesian updating} for a concrete example of measurement of momentum). They thus constitute the {\it instrumental} hidden variables of the model. Now, consider again a pair of independent preparations of two non-interacting subsystems with configuration $q=(q_A,q_B)$ and momentum $p=(p_A,p_B)$, so that it is described by a separable total wave function $\psi(q_A,q_B)=\psi_A(q_A)\psi_B(q_B)$. In this case, noting that the amplitude of the wave function is separable and the phase is decomposable, the distribution of the instrumental hidden variables determining measurement outcome associated with the independent preparations is given by 
\begin{eqnarray}
&&{\text P}_{\{S,\rho_{\tilde{p}}\}}(p,q,[\xi(t)])\nonumber\\
&=&\prod_{I=A,B}\delta\Big(p_I-\partial_{q_I}S_I-\frac{\xi}{2}\frac{\partial_{q_I}\rho_{\tilde{p}_I}}{\rho_{\tilde{p}_I}}\Big)\rho_{\tilde{p}_I}(q_I)\chi[\xi(t)]\nonumber\\
&\neq&{\text P}_{\{S_A,\rho_{\tilde{p}_A}\}}(p_A,q_A,[\xi(t)]){\text P}_{\{S_B,\rho_{\tilde{p}_B}\}}(p_B,q_B,[\xi(t)]),
\label{violation of preparation independence}
\end{eqnarray}
where $\chi[\xi(t)]$ is the probability density that $[\xi(t)]$ occurs. As explicitly shown above, due to the nonseparability (i.e. globalness) of $\xi$, thus the nonseparability of $[\xi(t)]$, the distribution of the instrumental hidden variables associated with independent preparations is not factorizable, violating preparation independence. Our epistemic interpretation of wave function therefore may not contradict PBR theorem, which states that under preparation independence, the wave function must be ontic or physical \cite{PBR theorem}.  

If instead $\xi$ is separable into two independent random variables, $[\xi(t)]$ must be also separable into two independent fluctuations, i.e., $[\xi(t)]=([\xi_A(t)],[\xi_B(t)])$ with $\chi[\xi(t)]=\chi_A[\xi_A(t)]\chi_B[\xi_B(t)]$. Inserting this into the second line in Eq. (\ref{violation of preparation independence}), it is clear that the distribution of the instrumental hidden variables associated with independent preparations becomes separable satisfying the preparation independence, i.e.: ${\text P}_{\{S,\rho_{\tilde{p}}\}}(p,q,[\xi(t)])={\text P}_{\{S_A,\rho_{\tilde{p}_A}\}}(p_A,q_A,[\xi_A(t)]){\text P}_{\{S_B,\rho_{\tilde{p}_B}\}}(p_B,q_B,[\xi_B(t)])$. In view of PBR theorem, the above analysis shows that for our epistemic interpretation to hold, the nonseparability of $\xi$ is indeed indispensible. This conclusion is consistent with the fact that, for separable $\xi$, the model does not lead to quantum mechanics for Hamiltonian with cross terms in momentum degrees of freedom, as e.g. arising in measurement interaction generating quantum entanglement (see Subsection \ref{Emergent quantum calculus} and Ref. \cite{Agung-Daniel model}). A closely related idea is reported in Ref. \cite{Emerson violation of preparation independence}. Note that the preparation independence is also regained in the classical limit, or in the formal limit $\xi\rightarrow 0$.  

\subsection{How the agent should update the estimation: Schr\"odinger equation and wave function collapse \label{Schroedinger equation and wave function collapse}}

To make the epistemic interpretation of wave function self-consistent, here we argue that the dynamics of the wave function, i.e., the unitary Schr\"odinger equation when there is no measurement, and nonunitary wave function collapse in measurement, also admit natural epistemic interpretations as normative rules to update the agent's estimation given information on the experimental settings. Suppose first that the agent does not make any selection of trajectories, hence, she/he does not have new relevant information for her/his estimation. In this case, the agent should rationally assume that her/his estimation must comply with conservation of trajectories which is manifested by the continuity equation. Furthermore, for the same reason, the agent should rationally also assume that her/his estimation respects the conservation of ensemble average energy (since the underlying momentum field is random, the energy of a single trajectory is in general not conserved).  As shown in Ref. \cite{Agung-Daniel model} and mentioned in Subsection \ref{Emergent quantum calculus}, these two conditions combined with the agent's epistemic decomposition of momentum field of Eq. (\ref{fundamental epistemic restriction}), imply that the time evolution of the wave function defined in Eq. (\ref{wave function}) | which represents the agent's estimation about her/his preparation | must satisfy the unitary Schr\"odinger equation. We emphasize that the two conditions, i.e., the conservation of trajectories and average energy, are not agent-independent objective dynamical constraints, like the principle of least action of classical mechanics. Rather, they should be seen as an application of a form of Bayesian inference taking into account the time symmetry of the statistical problem \cite{Jaynes principle of indifference}. Hence, they are epistemic subjective constraints, relative to the agent describing the system. As discussed below, other agent, having more detailed information about the system, must have instead a nonunitary time evolution for the wave function representing the updating of her/his estimation.    

Suppose instead that the agent makes a selection of certain subset of trajectories compatible with some macroscopically distinguishable values of a physical parameter, e.g., selecting those trajectories passing through one of the two arms in the Mach-Zehnder interferometer. In this case, since some (potential) trajectories of the original ensemble are no longer relevant for the agent's estimation, a rational agent must no longer impose the conservation of trajectories and average energy to update her/his estimation. Accordingly, the time evolution of wave function representing her/his estimation no longer follows the unitary Schr\"odinger equation \cite{Agung-Daniel model}. Moreover, the information obtained via the selection provides new evidence for the agent implying a Bayesian updating of the agent's weakly unbiased best estimation of the underlying momentum field. Finally, since the agent's estimation is represented by a wave function, the Bayesian updating of the agent's estimation is represented mathematically as a discontinuous change of wave function, or nonunitary wave function collapse. Of course, to be able to make a selection of trajectories, the system must be interacting with an apparatus of measurement in a specific way.  Unlike in classical mechanics, as discussed in Appendix \ref{An illustration of wave function collapse in measurement momentum as Bayesian updating}, this process of selection requires a specific disturbance to the underlying momentum field, while respecting the epistemic restriction of Eq. (\ref{fundamental epistemic restriction}) implying the MS errors trade-off of Eq. (\ref{trade off between MS error of position and momentum}) (or, the information trade-off  Eq. (\ref{the seed of quantum uncertainty})). 

The above subjective agent-dependent criterion, i.e., whether the agent has made a selection of trajectories compatible with some values of macroscopic parameter, or not, defines unambiguously whether or not there is a measurement. The time reversibility of the Schr\"odinger equation reflects the fact that there is no selection of trajectories, so that in this sense, there is no loss of information. On the other hand, the irreversibility of epistemic wave function collapse | which is necessary to have a consistent measurement |, reflects a loss of information due to selection of trajectories. As demonstrated in Appendix \ref{An illustration of wave function collapse in measurement momentum as Bayesian updating} with a concrete simple example for the measurement of momentum, the aim of a strong (i.e., standard, ideal) measurement is to manipulate the momentum field via measurement interaction allowing a selection of trajectories, so that for the selected ensemble of trajectories, the agent can make a sharp best estimation with vanishing estimation error. This epistemic requirement implies the standard projection postulate in quantum mechanics. Hence, measurement is a subjective agent-dependent phenomena with an active agent participation. Moreover, there is no infinite regress due to the ambiguity of the Heisenberg's cut (i.e., the shifty split), demarcating the system to be observed and the observer. The Heisenberg's cut is a subjective agent-dependent line reflecting a transfer of information via an act of selection of trajectories.  

Now consider two agents: one agent, Wigner, has no access to the selection of trajectories; and the other agent, Wigner's friend, has access to the selection of trajectories via measurement of position. In this case, according to Wigner's friend, within the above estimation scheme, after making the selection of trajectories, she should not impose the conservation of trajectories and average energy to update her estimation. Accordingly, the wave function representing her estimation about the system does not follow the Schr\"odinger equation. Rather, it must follow an epistemic wave function collapse onto one of the eigenfunctions of the quantum observable being observed, representing the Bayesian updating in light of new information obtained via the selection of trajectories compatible with the measurement outcome. By contrast, according to Wigner, since he has no access to the selection of trajectories (the only information that is available to Wigner is that the system is interacting with the measuring device with a specific interaction Hamiltonian, and observed by his friend), he should rationally impose the conservation of trajectories and average energy to update his estimation about the whole system-device-his friend. Accordingly, his wave function for describing his estimation about the whole system follow the unitary Schr\"odinger equation for measurement interaction, so that he has to represent his estimation with an entangled wave function between the system, the measuring apparatus, and state of mind of his friend. 

The two different descriptions of the system by Wigner and his friend are equally valid; i.e., they have different and incompatible information about the system, so that they must have different estimation and therefore must attribute different wave functions to their systems. The above analysis therefore again shows that wave function is indeed agent-dependent. Moreover, since measurement outcome is obtained via Bayesian reasoning, involving agent-dependent disturbance (see Appendix \ref{An illustration of wave function collapse in measurement momentum as Bayesian updating}) based on the agent-dependent estimation about the preparation represented by the wave function, it too is not an objective agent-independent fact, but has meaning only relative to the agent making the preparation and measurement. In this respect, similar to consistent histories \cite{Griffiths consistent histories}, relational quantum mechanics \cite{Rovelli relational QM}, and QBism \cite{Fuchs QBism}, the present epistemic model thus violates the assumption of consistency in an extension of Wigner's friend paradox devised by Frauchiger and Renner \cite{Frauchiger-Renner no go theorem}, blocking the various combination of statements by different agents. 

\section{Summary and discussion}

We have first started from a realist picture of microscopic world with an underlying momentum field $\tilde{p}(q;\xi)$, fluctuating randomly induced by a sub-quantum fluctuation of a global variable $\xi$ satisfying Eq. (\ref{Planck constant}). We then assumed an epistemic restriction that unlike in classical mechanics, the allowed probability distribution of positions $\rho_{\tilde{p}}(q)$ is fundamentally restricted by the form of the underlying momentum field $\tilde{p}(q;\xi)$. Now, suppose an agent wishes to estimate the momentum, given information on positions sampled from $\rho_{\tilde{p}}(q)$, under such an epistemic restriction. In this operational setting, we showed that quantum wave function emerges to represent the agent's classically consistent, weakly unbiased, ``best'' estimation of the momentum field with a specific estimation error. Hence, wave function is not an objective attribute of reality, but is epistemic and agent-dependent. Quantum uncertainty and complementarity between momentum and position finds its informational origin from the trade-off between the MS errors of simultaneous estimations of momentum field and mean position, with Gaussian wave functions represent the ``efficient'' simultaneous estimations, achieving the Cramer-Rao bounds of the associated MS errors. We then argued that unitary time evolution and wave function collapse in measurement are normative rules for an agent to rationally update her/his estimation given information on the experimental settings. We have thus ended up with an epistemic interpretation of quantum calculus, as a set of normative rules for an agent to predict the statistical results of measurement, employing a Bayesian reasoning, based on her/his best estimation about the preparation. 

There are three key conceptual ingredients in the above epistemic interpretation and reconstruction of quantum mechanics. First is the epistemic restriction, which makes explicit that the apparently innocuous Born's probability distribution of position, $\rho_{\tilde{p}}(q)=|\psi(q)|^2$, is actually implicitly parameterized by an underlying momentum field $\tilde{p}(q;\xi)$, thus in this sense contextual. We have shown that, by making explicit the underlying momentum fields (i.e., the contexts), the interference term in position space arising in coherence superposition of two wave functions can be explained without violating the usual law of total probability. The second conceptual ingredient is that the mathematical rules of quantum mechanics emerges in the operational setting of estimation, by an outside agent, about her/his preparation, based on which she/he rationally make predictions, via Bayesian reasoning, on the outcomes of measurements. The interpretation thus goes along the line of the Copenhagen Interpretation \cite{Bohr Q epistemic} and QBism \cite{Caves informational wave function,Fuchs QBism}, wherein the description of quantum mechanics is agent-dependent. Unlike those interpretations, however, we have derived the abstract rules of quantum calculus from a set of transparent epistemic or informational axioms, without assuming any quantum structures in Hilbert space. Third, we employ the unambiguous language of classical physics \cite{Bohr classical language}, from the outset, allowing a transparent quantum-classical transition and correspondence. 

Several important problems are left for future study. First, having conjectured that there is an underlying ontic dynamics, i.e., the dynamics of the random momentum field for an individual system, the most pressing problem concerns the explanation of the well-known no-go theorems, which conceptually limit such a conjecture. We have given a sketch on how the model may evade two recently devised no go theorems, that of PBR theorem via the violation of preparation independence \cite{PBR theorem}, and FR theorem via the violation of consistency assumption \cite{Frauchiger-Renner no go theorem}. Next, to complete the above epistemic reconstruction, and to better understand the puzzle that a slight modification of the Hilbert space quantum mechanics may lead to implausible implications stated in the Introduction, it is crucial to search for a set of transparent and plausible axioms which single out the specific functional form of the estimation error $\epsilon_p(q;\xi)$ of Eq. (\ref{estimation error}) \cite{Axioms for estimation error}. Related to this, since quantum mechanics emerges in the operational setting of best estimation, we should ask whether it is the ultimate (maximal) epistemic theory, or, whether it is possible to go beyond the epistemic description of quantum mechanics, into the sub-quantum ontic fluctuation. In particular, it is intriguing to ask if we can or cannot experimentally probe the single shot estimation error $\epsilon_p(q;\xi)$ of Eq. (\ref{efficient estimation of momentum and random error}). Further, one wonders how quantum mechanics, being shown as calculus for estimation under epistemic restriction, be reconciled with relativity.  

\begin{acknowledgments}

The present work is partially supported by the John Templeton Foundation (project ID 43297). The opinions expressed in this publication do not necessarily reflect the views of the John Templeton Foundation. It is also in part supported by the Indonesia Ministry of Research, Technology, and Higher Education (MRTHE) through PDUPT research scheme (Contract No. 2/E1/KP.PTNBH/2019) and the Institut Teknologi Bandung (ITB) WCU Program. I would like to thank Daniel Rohrlich for insightful and stimulating discussion, and Michael Hall and the anonymous Referees for constructive criticism and recommendation, and Hermawan K. Dipojono for his generous support and useful discussion.

\end{acknowledgments}

\appendix

\section{Derivation of Cram\'er-Rao inequality for the estimation of momentum of Eq. (\ref{Cramer-Rao bound for momentum estimation})\label{derivation of Cramer-Rao bound}}

Suppose an agent wants to estimate the momentum field $\tilde{p}(q;\xi)$ arising in her/his preparation arrangement, based on information on position $q$ sampled from $\rho_{\tilde{p}}(q)$ parameterized by the momentum field $\tilde{p}(q;\xi)$ via a fundamental epistemic restriction. Assume that the estimator $\partial_qS (q)$ is in a sense `weakly' unbiased so that we impose:
\begin{eqnarray}
\int{\rm d}q~\partial_qS (q)\rho_{\tilde{p}}(q)=\int{\rm d}q~\tilde{p}(q;\xi)\rho_{\tilde{p}}(q). 
\label{unbiased condition}
\end{eqnarray} 
The above condition is weaker than the usual unbiased condition which demands that the average of the estimator is equal to the parameter being estimated \cite{Papoulis and Pillai book on probability and statistics,Cover-Thomas book}. 

First, we take the derivative of both sides of Eq. (\ref{unbiased condition}) with respect to $\tilde{p}$ to get 
\begin{eqnarray}
\int{\rm d}q\big(\tilde{p}(q;\xi)-\partial_qS (q)\big)\frac{\partial\rho_{\tilde{p}}(q)}{\partial\tilde{p}}=-\int{\rm d}q\rho_{\tilde{p}}(q)=-1, 
\label{unbiased condition plus regularity}
\end{eqnarray}
where we have used a regularity condition \cite{Papoulis and Pillai book on probability and statistics}: $\frac{\partial}{\partial \tilde{p}}\int{\rm d}q(\cdot)=\int{\rm d}q\frac{\partial}{\partial\tilde{p}}(\cdot)$. Then, we rewrite Eq. (\ref{unbiased condition plus regularity}) as 
\begin{eqnarray}
\int{\rm d}q\Big(\big(\tilde{p}(q;\xi)-\partial_qS (q)\big)\sqrt{\rho_{\tilde{p}}(q)}\Big)\Big(\sqrt{\rho_{\tilde{p}}(q)}\frac{1}{\rho_{\tilde{p}}(q)}\frac{\partial \rho_{\tilde{p}}(q)}{\partial\tilde{p}}\Big)=-1.
\label{Cramer-Rao bound -1}
\end{eqnarray}
Applying the Cauchy-Schwarz inequality, we thus obtain 
\begin{eqnarray}
\int{\rm d}q\big(\tilde{p}(q;\xi)-\partial_qS(q)\big)^2\rho_{\tilde{p}}(q)\int{\rm d}q\Big(\frac{1}{\rho_{\tilde{p}}(q)}\frac{\partial\rho_{\tilde{p}}(q)}{\partial\tilde{p}}\Big)^2\rho_{\tilde{p}}(q)\ge 1.
\label{Cramer-Rao bound 0}
\end{eqnarray}
For each $\xi$, the first integral on the left hand side is just the MS error of the estimation of the momentum field $\tilde{p}(q;\xi)$ with the weakly unbiased estimator $\partial_qS(q)$, so that we have the well-known Cram\'er-Rao inequality \cite{Papoulis and Pillai book on probability and statistics,Cover-Thomas book}:
\begin{eqnarray}
\int{\rm d}q\big(\tilde{p}(q;\xi)-\partial_qS (q)\big)^2\rho_{\tilde{p}}(q)\ge\frac{1}{J_p},
\label{Cramer-Rao bound appendix}
\end{eqnarray}
where $J_p$ is the Fisher information about $\tilde{p}$ contained in $\rho_{\tilde{p}}(q)$ defined as 
\begin{eqnarray}
J_p\doteq\int{\rm d}q\big(\partial_{\tilde{p}}\ln\rho_{\tilde{p}}(q)\big)^2\rho_{\tilde{p}}(q). 
\label{Fisher information for momentum}
\end{eqnarray}
The weakly unbiased estimator $\partial_qS(q)$ is called an ``efficient'' estimator if it is optimal saturating the Cram\'er-Rao inequality of Eq. (\ref{Cramer-Rao bound appendix}), i.e.: 
\begin{eqnarray}
\int{\rm d}q\big(\tilde{p}(q;\xi)-\partial_qS (q)\big)^2\rho_{\tilde{p}}(q)=\frac{1}{J_p}. 
\label{efficient estimator}
\end{eqnarray}
From Eq. (\ref{Cramer-Rao bound -1}), this is achieved when the estimator $\partial_qS(q)$ and the parameter being estimated $\tilde{p}(q;\xi)$ satisfies the following condition \cite{Papoulis and Pillai book on probability and statistics}:
\begin{eqnarray}
\tilde{p}(q;\xi)=\partial_qS (q)+a(\xi)\partial_{\tilde{p}} \ln\rho_{\tilde{p}}(q).
\label{Cramer-Rao optimal condition}
\end{eqnarray}
where $a$ is independent of $q$.  

\section{Efficient estimation of mean position and Gaussian distribution of error \label{Efficient estimation of mean position and Gaussian distribution of error}}

Suppose that the estimation of the mean position $q_o$ parameterizing $\rho_{\tilde{p}}(q)$ with the unbiased estimator $q$ is efficient, so that the Cram\'er-Rao inequality of Eq. (\ref{Cramer-Rao inequality for position}) is saturated, i.e.: $\mathcal{E}_q^2\doteq\int{\rm d}q(q-q_o)^2\rho_{\tilde{p}}(q)=1/J_q$.  In this case, the estimator $q$ and the parameter being estimated $q_o$ must satisfy the following condition \cite{Papoulis and Pillai book on probability and statistics}:
\begin{eqnarray}
q_o=q+b\frac{\partial_q\rho_{\tilde{p}}}{\rho_{\tilde{p}}},
\end{eqnarray}
where $b>0$. Integrating the above differential equation, denoting $b=\sigma_q^2$, the distribution of error of efficient estimation of $q_o$ with the estimator $q$ must therefore be given by a Gaussian: 
\begin{eqnarray}
\rho_{\tilde{p}}(q)=\frac{1}{\sqrt{2\pi\sigma_q^2}}e^{-\frac{(q-q_o)^2}{2\sigma_q^2}}, 
\label{Gaussian distribution of error}
\end{eqnarray}
as claimed in the main text. 

\section{Decomposition of quantum variance of momentum into the accuracy and imprecision of best estimation of momentum \label{Decomposition of quantum variance of momentum into the accuracy and impression of best estimation of momentum}}

Writing the wave function in polar form as in Eq. (\ref{wave function}), and computing the quantum variance $\sigma_{\hat p}^2$ of the momentum operator $\hat{p}$, one directly obtains 
\begin{eqnarray}
\sigma_{\hat p}^2&\doteq&\braket{\psi|\hat{p}^2|\psi}-(\braket{\psi|\hat{p}|\psi})^2\nonumber\\
&=&-\hbar^2\int{\rm d}q\frac{\partial_q^2\sqrt{\rho_{\tilde{p}}}}{\sqrt{\rho_{\tilde{p}}}}\rho_{\tilde{p}}(q)+\int{\rm d}q(\partial_qS )^2\rho_{\tilde{p}}(q)-\Big(\int{\rm d}q\partial_qS \rho_{\tilde{p}}(q)\Big)^2\nonumber\\
&=&\frac{\hbar^2}{4}\int{\rm d}q\Big(\frac{\partial_q\rho_{\tilde{p}}}{\rho_{\tilde{p}}}\Big)^2\rho_{\tilde{p}}(q)+\int{\rm d}q\Big(\partial_qS -\int{\rm d}q'\partial_{q'}S \rho_{\tilde{p}}(q')\Big)^2\rho_{\tilde{p}}(q).
\label{quantum uncertainty momentum}
\end{eqnarray}
Here to get the first term in the third line, we have first used a mathematical identity 
\begin{eqnarray}
-\frac{\partial_q^2\sqrt{\rho_{\tilde{p}}}}{\sqrt{\rho_{\tilde{p}}}}=\frac{1}{4}\Big(\frac{\partial_q\rho_{\tilde{p}}}{\rho_{\tilde{p}}}\Big)^2-\frac{1}{2}\frac{\partial_q^2\rho_{\tilde{p}}}{\rho_{\tilde{p}}},
\label{quantum identity}
\end{eqnarray}
to replace the first term of the second line, followed by integrating the second term on the right hand side of Eq. (\ref{quantum identity}) after being multiplied by $\rho_{\tilde{p}}(q)$, applying the fundamental theorem of calculus and discarding the surface term assuming that the first derivative of $\rho_{\tilde{p}}(q)$ is vanishing at the boundary. Further, the last two terms of the second line in Eq. (\ref{quantum uncertainty momentum}) is rewritten as $\int{\rm d}q(\partial_qS )^2\rho_{\tilde{p}}(q)-\big(\int{\rm d}q\partial_qS \rho_{\tilde{p}}(q)\big)^2=\int{\rm d}q\big(\partial_qS -\int{\rm d}q'\partial_{q'}S\rho_{\tilde{p}}(q')\big)^2\rho_{\tilde{p}}(q)\doteq\Delta_p^2$, which is just the dispersion of the estimator $\partial_qS(q)$. We thus obtain $\sigma_{\hat{p}}^2=\mathcal{E}_p^2+\Delta_p^2$ as claimed in the main text. 

\section{Gaussian wave function describes efficient estimation of momentum\label{Gaussian wave function describes an efficient estimation of momentum}}

We show that when the preparation is given by Gaussian wave function of Eq. (\ref{Gaussian wave function}), the estimation of the momentum field $\tilde{p}$ by the estimator $\partial_qS(q)=p_o$ based on the decomposition of Eq. (\ref{fundamental epistemic restriction}) is also efficient saturating the associated Cram\'er-Rao inequality of Eq. (\ref{Cramer-Rao bound appendix}). First, since $\rho_{\tilde{p}}(q)$ is given by Gaussian distribution of Eq. (\ref{Gaussian distribution of error}), noting Eq. (\ref{efficient estimation of momentum and random error}), the estimation error of momentum is linear in position given by 
\begin{eqnarray}
\epsilon_p(q;\xi)=\frac{\xi}{2}\frac{\partial_q\rho_{\tilde{p}}}{\rho_{\tilde{p}}}=-\frac{\xi}{2}\frac{q-q_o}{\sigma_q^2}. 
\end{eqnarray}
Inserting into Eq. (\ref{fundamental epistemic restriction}), and noting that for Gaussian wave function $\partial_qS(q)=p_o$ is independent of $q$, we have 
\begin{eqnarray}
a\doteq\frac{\xi}{2}\frac{\partial\tilde{p}}{\partial q}=-\frac{\xi^2}{4\sigma_q^2}. 
\label{constant proportionality for efficient Gaussian}
\end{eqnarray}
Using $a$ defined above, we can then rewrite Eq. (\ref{fundamental epistemic restriction}) associated with Gaussian wave function as 
\begin{eqnarray}
\tilde{p}(q;\xi)=p_o+a\frac{\partial_{\tilde{p}}\rho_{\tilde{p}}}{\rho_{\tilde{p}}}. 
\end{eqnarray}
Since $a$ is independent of $q$, as shown at the end of Appendix \ref{derivation of Cramer-Rao bound}, the above equation can be seen as the condition for an efficient estimation of momentum with the estimator $p_o$.       

\section{Estimation of momentum field associated with superposition of two wave functions\label{Efficient estimation of momentum and its error for the superposition of two wave functions}}

Consider the superposition of two wave functions as follows:
\begin{eqnarray}
\psi(q)\sim\psi_1(q)+\psi_2(q),~~{\rm where} ~~\psi_j=\sqrt{\rho_{\tilde{p}_j}(q)}e^{\frac{i}{\hbar}S_j(q)}, ~~j=1,2.
\label{superposition of two wave functions}
\end{eqnarray}
Inserting into Eq. (\ref{efficient estimation of momentum and random error}), the weakly unbiased best estimate of the momentum given information on position, and the associated estimation error, read
\begin{eqnarray}
\overline{p}(q)=\partial_qS &=&\frac{1}{\rho_{\tilde{p}_1}+\rho_{\tilde{p}_2}+2\sqrt{\rho_{\tilde{p}_1}\rho_{\tilde{p}_2}}\cos(\frac{S_1 -S_2 }{\hbar})}\nonumber\\
&\times&\Big(\rho_{\tilde{p}_1}\partial_qS_1 +\rho_{\tilde{p}_2}\partial_qS_2 \nonumber\\
&+&\sqrt{\rho_{\tilde{p}_1}\rho_{\tilde{p}_2}}\cos\Big(\frac{S_1 -S_2 }{\hbar}\Big)(\partial_qS_1 +\partial_qS_2 )\nonumber\\
&-&\hbar(\sqrt{\rho_{\tilde{p}_1}}\partial_q\sqrt{\rho_{\tilde{p}_2}}-\sqrt{\rho_{\tilde{p}_2}}\partial_q\sqrt{\rho_{\tilde{p}_1}})\sin\Big(\frac{S_1 -S_2 }{\hbar}\Big)\Big), \nonumber\\
\epsilon_p(q;\xi)&=&\frac{\xi}{2}\frac{1}{\rho_{\tilde{p}_1}+\rho_{\tilde{p}_2}+2\sqrt{\rho_{\tilde{p}_1}\rho_{\tilde{p}_2}}\cos(\frac{S_1 -S_2 }{\hbar})}\nonumber\\
&\times&\Big(\partial_q\rho_{\tilde{p}_1}+\partial_q\rho_{\tilde{p}_2}+(2\sqrt{\rho_{\tilde{p}_1}}\partial_q\sqrt{\rho_{\tilde{p}_2}}+2\sqrt{\rho_{\tilde{p}_2}}\partial_q\sqrt{\rho_{\tilde{p}_1}})\cos\Big(\frac{S_1 -S_2 }{\hbar}\Big)\nonumber\\
&-&\frac{2}{\hbar}\sqrt{\rho_{\tilde{p}_1}\rho_{\tilde{p}_2}}\sin\Big(\frac{S_1 -S_2 }{\hbar}\Big)(\partial_qS_1 -\partial_qS_2 )\Big). 
\label{general formula for the efficient estimation and errors for superposition}
\end{eqnarray}

One can thus see that when $\psi_1$ and $\psi_2$ are not overlapping, namely when $\rho_{\tilde{p}_1}(q)\rho_{\tilde{p}_2}(q)=0$ for all $q$, the above two equations reduce into 
\begin{eqnarray}
\overline{p}(q)&=&\frac{\rho_{\tilde{p}_1}\partial_qS_1 +\rho_{\tilde{p}_2}\partial_qS_2 }{\rho_{\tilde{p}_1}+\rho_{\tilde{p}_2}}=\mu_1\overline{p}_1(q)+\mu_2\overline{p}_2(q),\nonumber\\
\epsilon_p(q;\xi)&=&\frac{\xi}{2}\frac{\partial_q\rho_{\tilde{p}_1}}{\rho_{\tilde{p}_1}+\rho_{\tilde{p}_2}}+\frac{\xi}{2}\frac{\partial_q\rho_{\tilde{p}_2}}{\rho_{\tilde{p}_1}+\rho_{\tilde{p}_2}}=\mu_1\epsilon_{p_1}(q;\xi)+\mu_2\epsilon_{p_2}(q;\xi),
\label{efficient estimation and error for non-overlapping superposition 0}
\end{eqnarray}
where $\overline{p}_j(q)=\partial_qS_j(q)$ and $\epsilon_{p_j}(q;\xi)=\frac{\xi}{2}\partial_q\ln\rho_{\tilde{p}_j}(q)$ are the weakly unbiased best estimate of momentum field and its estimation error associated with preparation represented by $\psi_j$, and 
\begin{eqnarray}
\mu_j(q)\doteq\frac{\rho_{\tilde{p}_j}(q)}{\rho_{\tilde{p}_1(q)}+\rho_{\tilde{p}_2}(q)},
\end{eqnarray}
$j=1,2$. Denoting the support of $\psi_j(q)$ as $\Lambda_j$, $j=1,2$, since $\Lambda_1\cap\Lambda_2=\emptyset$, then for $q\in\Lambda_j$ we have $\mu_k(q)=\delta_{jk}$, $k=1,2$. Noting this, for $q\in\Lambda_j$, the two equations in (\ref{efficient estimation and error for non-overlapping superposition 0}) can be written as 
\begin{eqnarray}
\overline{p}(q)&=&\overline{p}_j(q),\nonumber\\
\epsilon_p(q;\xi)&=&\epsilon_{p_j}(q;\xi). 
\label{best estimate and error for nonoverlapping superposition}
\end{eqnarray}
For $q\in\Lambda_j$, $j=1,2$, one therefore has 
\begin{eqnarray}
\tilde{p}(q;\xi)&=&\overline{p}(q)+\epsilon_p(q;\xi)\nonumber\\
&=&\overline{p}_j(q)+\epsilon_{p_j}(q;\xi)=\tilde{p}_j(q;\xi).
\end{eqnarray}

\section{Wave function collapse in measurement represents a Bayesian updating of the agent's estimation: a simple example \label{An illustration of wave function collapse in measurement momentum as Bayesian updating}}

We give a concrete simple example of an epistemic wave function collapse in measurement due to a Bayesian updating of the agent's estimation via a selection of trajectories. Consider an interaction between a system with a spatial coordinate $q_A$ and a device for momentum measurement with the coordinate of its pointer $q_B$, via a measurement interaction classical Hamiltonian $H=gp_Ap_B$, where $p_A$ and $p_B$ are respectively the momentum of the system and the pointer of the measuring device, and $g$ is the interaction strength. We assume that the measurement interaction is impulsive so that the effects of individual free Hamiltonians can be neglected. Suppose the initial wave function of the system-device at time $t=t_0$ is separable as $\psi(q_A,q_B,t_0)=\psi_A(q_A,t_0)\psi_B(q_B,t_0)$, representing the agent's initial estimation that the system and the device are prepared independently. Further, for the purpose of illustration, assume that the initial wave function of the system takes the simplest yet nontrivial form: $\psi_A(q_A,t_0)\sim e^{\frac{i}{\hbar}p_oq_A}+e^{-\frac{i}{\hbar}p_oq_A}\sim\cos(p_oq_A/\hbar)$; extension to a general form of initial wave function is straightforward. Hence, as per Eq. (\ref{wave function}), initially at $t=t_0$, since $S_A(q_A,t_0)=0$ and $\rho_{\tilde{p}_A}(q_A,t_0)\sim\cos^2(p_oq_A/\hbar)$, using Eq. (\ref{efficient estimation of momentum and random error}), the agent's best estimate of the momentum of the system given $q_A$ is vanishing, $\overline{p}_A(q_A,t_0)=\partial_{q_A}S_A(q_A,t_0)=0$, with a non-vanishing random estimation error $\epsilon_{p_A}(q_A,t_0;\xi)=\frac{\xi}{2}\partial_{q_A}\ln\rho_{\tilde{p}_A}(q_A,t_0)=-\frac{\xi}{\hbar}p_o\tan(p_oq_A/\hbar)$ (with MS error: $\mathcal{E}_{p_A}^2=p_o^2$). As we shall see, the aim of a strong (i.e., ideal or standard) measurement of the system momentum is to manipulate the momentum field of the system-device so that at the end of the process, the agent can make a sharp, weakly unbiased, best estimate of the momentum of the system given information on the position of the device pointer, with a vanishing estimation error. This epistemic requirement, as we argue below, implies the Dirac-von Neumann projection postulate that an immediate repetition of measurement must give the same outcome. 

To do this, the agent first lets the system and device interact without any selection of trajectories so that, she/he should rationally impose the conservation of trajectories and average energy to her/his estimation. As shown in Ref. \cite{Agung-Daniel model}, in this case, the associated wave function representing her/his estimation evolves following the Schr\"odinger equation with a measurement interaction quantum Hamiltonian $\hat{H}=g\hat{p}_A\hat{p}_B$. Solving the Schr\"odinger equation with the above separable initial wave function, because of the measurement interaction, the wave function at time $t=T$ becomes entangled: $\psi(q_A,q_B,T)\sim e^{\frac{i}{\hbar}p_oq_A}\psi_B(q_B-gp_oT)+e^{-\frac{i}{\hbar}p_oq_A}\psi_B(q_B+gp_oT)$. One can see that the wave function of the device splits into two wave packets, i.e., $\psi_B(q_B-gp_oT)$ and $\psi_B(q_B+gp_oT)$, respectively obtained by shifting the initial wave function of the device $\psi_B(q_B)$ a finite amount $gp_oT$ and $-gp_oT$, so that it correlates with each term of the superposition in the initial wave function of the system. 

Next, assume that the shifted wave functions of the device, i.e., $\psi_B(q_B-gp_oT)$ and $\psi_B(q_B+gp_oT)$, are not overlapping, so that they are distinguishable via the knowledge of $q_B$. To achieve such separation, we can choose sufficiently strong interaction $g$ or start with sufficiently narrow initial wave function  $\psi_B(q)$ of the device. Suppose further that the agent makes a selection of trajectories of the device pointer by recording $q_B$ at time $T$. Assume that $q_B$ at time $t=T$ enters the support of $\psi_B(q_B-gp_oT)$. Due to the selection of trajectories, the agent should rationally assume that the conservation of trajectories and average energy are no longer respected. The new information obtained in the selection (i.e., that $q_B$ at time $t=T$ enters the support of $\psi_B(q_B-gp_oT)$) must rationally sharpen the agent's estimation about the system-device, so that the associated wave function representing the agent's estimation must also transform, discontinuously and nonunitarily, as: $\psi(q_A,q_B,T)\sim e^{\frac{i}{\hbar}p_oq_A}\psi_B(q_B-gp_oT)+e^{-\frac{i}{\hbar}p_oq_A}\psi_B(q_B+gp_oT)$ (prior to selection) $\mapsto$ $\psi(q_A,q_B,T)\sim e^{\frac{i}{\hbar}p_oq_A}\psi_B(q_B-gp_oT)$ (after the selection). Accordingly, there is an effective wave function collapse of the system due to the updating of the agent's estimation about the system as: $\psi_A(q_A,t_0)\sim e^{\frac{i}{\hbar}p_oq_A}+e^{-\frac{i}{\hbar}p_oq_A}$ (prior to measurement interaction) $\mapsto \psi_A(q_A,T)\sim e^{\frac{i}{\hbar}p_oq_A}$ (after the measurement). Note importantly that, all the above processes involving the wave function occur not in physical space, but in the agent's rational mind. 

Next, from Eq. (\ref{efficient estimation of momentum and random error}), since after the measurement interaction at time $t=T$ we have $S_A(q_A,T)=p_oq_A$, and $\rho_{\tilde{p}_A}(q_A,T)$ is spatially uniform, the agent's best estimate of the momentum of the system given $q_B$ becomes $\overline{p}_A(T)=\partial_{q_A}S_A(T)=p_o$, with vanishing estimation error $\epsilon_{p_A}(T)=\frac{\xi}{2}\partial_{q_A}\ln\rho_{\tilde{p}_A}(q_A,T)=0$, as expected. Exactly the same process occurs when $q_B$ at time $t=T$ enters instead the support of $\psi_B(q_B+gp_oT)$, in case of which, the agent's best estimate of momentum after the measurement is $\overline{p}_A(T)=-p_o$, with vanishing estimation error $\epsilon_{p_A}(T)=0$. It is clear in the above mechanism that repeating the measurement immediately after the first measurement, since there is no new information, the agent's best estimate should yield the same outcome with vanishing estimation error. One can see that these sharp estimates of the system momentum at the end of measurement, i.e. $\overline{p}_A(T)=\pm p_o$, are exactly equal to the outcomes of quantum measurement of $\hat{p}$ over the initial wave function of the system $\psi_A(q_A,t_0)\sim e^{\frac{i}{\hbar}p_oq_A}+e^{-\frac{i}{\hbar}p_oq_A}$. Finally, we note that since the underlying momentum field is random due to the fluctuation of $\xi$, whether $q_B$ at time $t=T$ enters the support of $\psi_B(q_B-gp_oT)$ or $\psi_B(q_B+gp_oT)$ must occur randomly. Accordingly, the agent's best estimate of the momentum in each measurement repetition also fluctuates randomly $\overline{p}_A(T)=\pm p_o$. 

The above scheme of momentum measurement explicitly shows that a single measurement outcome | in the above example: $\pm p_o$ | does not in general reveal the objective value of momentum $\tilde{p}_A$ prior to measurement. Rather, as in the Copenhagen interpretation, a measurement is, in a sense, a way of preparing the system in some final state, to sharpen the agent's estimation. Hence it creates new value via the active participation of the agent. The measurement interaction changes the initial momentum field, allowing the agent to select a specific ensemble of trajectories compatible with a specific macroscopically distinguishable value of a parameter of the measuring device, so that for the selected ensemble she/he can make a sharp estimation of system momentum with vanishing estimation error. There is thus a Bayesian mapping, i.e., the sharpening of the agent's estimation about the momentum of the system: from $\overline{p}_A(t_0)=0$ (prior to measurement) with finite MS error, $\mathcal{E}_{p_A}^2(t_0)=p_o^2>0$, to $\overline{p}_A(T)=\pm p_o$ (after the measurement) with vanishing MS error, $\mathcal{E}_{p_A}^2(T)=0$. We note that after the momentum measurement, since the final wave function is a plane wave, the agent's unbiased estimate of the mean position becomes infinitely poor, i.e., $\mathcal{E}_{q_A}^2(T)=\infty$, in accord with the trade-off of the MS errors of Eq. (\ref{trade off between MS error of position and momentum}). Further, the mapping occurs randomly with a probability that can be shown to reproduce the prediction of quantum mechanics as prescribed by the Born's rule (see Section Methods in Ref. \cite{Agung-Daniel model}). The above results for the measurement of momentum can be straightforwardly extended to general initial wave function of the system, and also apply to the measurement of angular momentum with discrete possible measurement outcomes.


\begin{thebibliography}{10}   

\bibitem{Bohm pilot-wave theory} D. Bohm, Phys. Rev. 85, 166 (1952); D. Bohm and B. J. Hiley, The Undivided Universe: an Ontological Interpretation of Quantum Theory, Routledge, London, (1993).
\bibitem{Everett many worlds} H. Everett, Rev.Mod. Phys. 29, 454 (1957). 
\bibitem{Wheeler many worlds} J. A. Wheeler, Rev. Mod. Phys. 29, 463 (1957).
\bibitem{deWitt many worlds} B. S. DeWitt, Physics Today 23, 30 (1970).
\bibitem{GRW theory} G. C. Ghirardi, A. Rimini, and T. Weber, Phys. Rev. D 34, 470 (1986). 

\bibitem{Bohr Q epistemic} N. Bohr, Atomic Physics and Human Knowledge, Wiley, New York, (1958); M. Jammer, The Philosophy of Quantum Mechanics:
The Interpretations of Quantum Mechanics in Historical Perspective, Wiley-Interscience, New York (1974).
\bibitem{Ballentine statistical interpretation} L. E. Ballentine, Rev. Mod. Phys. 42, 358 (1970). 
\bibitem{Griffiths consistent histories} R. B. Griffiths, J. Stat. Phys. 36, 219 (1984); ``The Consistent Histories Approach to Quantum Mechanics'', Stanford Encyclopedia of Philosophy (2019), https://plato.stanford.edu/entries/qm-consistent-histories/. 
\bibitem{Omnes consistent histories} R. Omnes, J. Stat. Phys. 53, 893 (1988). 
\bibitem{Gell-Mann and Hartle decoherent histories} M. Gell-Mann and J. Hartle, in W. H. Zurek (Ed.), Complexity, Entropy, and the Physics of Information, SFI Studies in the Sciences of Complexity, vol III, Addison Wesley (1990). 
\bibitem{Rovelli relational QM} C. Rovelli, Int. J. Theor. Phys. 35, 1637 (1996).
\bibitem{Zeilinger axioms} A. Zeilinger, Found. Phys. 29, 631 (1999).
\bibitem{Caves informational wave function} C. M. Caves, C. A. Fuchs, and R. Schack, Phys. Rev. A 65, 022305 (2002).
\bibitem{Fuchs QBism} C. A. Fuchs and R. Schack, Rev. Mod. Phys. 85 (4), 1693 (2013).  

\bibitem{Jaynes ontic-epistemic} E. T. Jaynes, Foundations of Radiation Theory and Quantum Electrodynamics, Plenum, New York (1980).
\bibitem{Harrigan ontic-epistemic} N. Harrigan and R. W. Spekkens, Found. Phys. 40, 125 (2010).

\bibitem{Zurek quantum-classical correspondence} W. H. Zurek, Phys. Scripta T 76, 186 (1998). 
\bibitem{Emerson PhD thesis} J. V. Emerson, Quantum Chaos and Quantum-Classical Correspondence, PhD dissertation, Simon Fraser University (2001).

\bibitem{Wheeler: why} J. A. Wheeler, in W.H. Zurek (Ed.), Complexity, Entropy and the Physics of Information, Westview Press (1990).
\bibitem{Popescu-Rohrlich axioms} S. Popescu and D. Rohrlich, Found. Phys. 24, 379 (1994).

\bibitem{Gisin nonlinearity - signaling} N. Gisin, Physics Letters A 143, 1 (1990); Helvetica Physica Acta 62, 363 (1989).
\bibitem{Polchinski nonlinearity - signaling} J. Polchinski, Phys. Rev. Lett. 66, 397 (1991). 
\bibitem{Lee nonHermitian - signaling} Y. C. Lee, M. H. Hsieh, S. T. Flammia, and R. K. Lee, Phys. Rev. Lett. 112, 13040 (2014).  
\bibitem{Peres nonlinearity violates 2nd law} A. Peres, Phys. Rev. Lett. 63, 1114 (1989). 
\bibitem{Hanggi UR second law} E. H\"anggi and S. Wehner, Nat. Commun. 4, 1670 (2013). 

\bibitem{Caves-Fuchs axioms} C. M. Caves and C. A. Fuchs, Quantum information: how much information in a state vector?, arXiv:quant-ph/9601025, (1996).
\bibitem{Hardy axioms} L. Hardy, Quantum theory from five reasonable axioms, arXiv:quant-ph/0101012, (2001).
\bibitem{Fuchs axioms} C. A. Fuchs, Quantum foundations in the light of quantum information, arXiv:quant-ph/0105039, (2001).
\bibitem{Spekkens toy model with epistemic restriction} R. W. Spekkens, Phys. Rev. A 75, 032110 (2007). 
\bibitem{Barrett axioms} J. Barrett, Phys. Rev. A 75, 032304 (2007).
\bibitem{Dakic-Brukner axioms} B. Dakic and C. Brukner, in H. Halvorson (Ed.), Deep Beauty: Understanding the Quantum World through Mathematical Innovation, Cambridge University Press, Cambridge (2011); arXiv:quant-ph/0911.0695, (2009).
\bibitem{Paterek axioms} T. Paterek, B. Dakic, and C. Brukner, New J. Phys. 12, 053037 (2010).
\bibitem{Chiribella axioms} G. Chiribella, G. M. D'Ariano, and P. Perinotti, Phys. Rev. A 84, 012311 (2011).
\bibitem{Masanes axioms} L. Masanes and M. P. M\"uller, New J. Phys. 13, 063001 (2011).
\bibitem{de la Torre axioms} G. de la Torre, L. Masanes, A. J. Short, and M. P. M\"uller, Phys. Rev. Lett. 109, 090403 (2012).
\bibitem{Bartlett reconstruction of Gaussian QM with epistemic restriction} S. D. Bartlett, T. Rudolph, and R. W. Spekkens, Phys. Rev. A 86, 012103 (2012).
\bibitem{Spekkens quasi-quantization} R. W. Spekkens, ``Quasi-Quantization: Classical Statistical Theories with an Epistemic Restriction," Quantum Theory: Informational Foundations and Foils pp. 83-138; arXiv:1409.5041.
\bibitem{Frieden model} B. R. Frieden, Am. J. Phys. 57, 1004 (1989).
\bibitem{Reginatto estimation scheme} M. Reginatto, Phys. Rev. A 58, 1775 (1998). 
\bibitem{Hall-Reginatto model} M. J. W. Hall and M. Reginatto, J. Phys. A: Math. Gen. 35, 3289 (2002).
\bibitem{deRaedt model} H. De Raedt, M. I. Katsnelson, and K. Michielsen, Annals of Physics 347, 45 (2014).
\bibitem{Nelson stochastic mechanics} E. Nelson, Phys. Rev. 150, 1079 (1966).
\bibitem{Markopolou-Smolin quantum from cosmos} F. Markopoulou and L. Smolin, Phys. Rev. D 70, 124029 (2004).
\bibitem{Smolin quantum from cosmos} L. Smolin, Could quantum mechanics be an approximation to another theory?, arXiv:quant-ph/0609109v1, (2006).
\bibitem{Goyal information geometry} P. Goyal, New Journal of Physics 12, 023012 (2010).
\bibitem{Caticha quantum from MEP} A. Caticha, J. Phys. A 44, 225303 (2011). 
\bibitem{Hohn quantum from question} P. A. H\"ohn and C. S. P. Wever, Phys. Rev. A 95, 012102 (2017). 
\bibitem{Selby quantum from diagram} J. H. Selby, C. M. Scandolo and B. Coecke, Reconstructing quantum theory from diagrammatic postulates, arXiv:1802.00367v2, (2018). 

\bibitem{Agung-Daniel model} A. Budiyono and D. Rohrlich, Nat. Comms. 8, 1306 (2017). 

\bibitem{Hardy epistricted model} L. Hardy, Disentangling nonlocality and teleportation, arXiv:quant-ph/9906123, (1999).

\bibitem{Papoulis and Pillai book on probability and statistics} A. Papoulis and S. U. Pillai, Probability, Random Variable and Stochastic Processes, McGraw-Hill, Singapore (2002). 
\bibitem{Cover-Thomas book} T. M. Cover and J. A. Thomas, Elements of Information Theory, 2-nd Edition, John Wiley and Sons, New Jersey (2006). 

\bibitem{Wigner's paradox} E. P. Wigner, in Symmetries and Reflections, Indiana University Press (1967).

\bibitem{Hall exact uncertainty and best estimation} M. J. W. Hall, Phys. Rev. A 64, 052103 (2001).

\bibitem{Deutch Born's rule from decision} D. Deutch, Proc. R. Soc. Lond. A 455, 3129 (1999).  

\bibitem{Rund book: Hamilton-Jacobi formalism} H. Rund, The Hamilton-Jacobi Theory in the Calculus of Variations: Its Role in Mathematics and Physics, Van Nostrand, London (1966).

\bibitem{Bell book} J. S. Bell, Speakable and Unspeakable in Quantum Mechanics, Cambridge University Press, Cambridge (1987). 
\bibitem{Popper book} K. Popper, Quantum Theory and The Schism in Physics, Routledge, New York, 1992. 

\bibitem{Agung ERPS representation} A. Budiyono, Phys. Rev. A 100, 032125 (2019). 

\bibitem{Bub pre-dynamics probabilistic constraint} J. Bub, Studies in History and Philosophy of Science Part B, in press (2018), arXiv:quant-ph/1804.03267. 

\bibitem{PBR theorem} M. F. Pusey, J. Barrett, and T. Rudolph, Nat. Phys. 8, 475 (2012).

\bibitem{Dirac book} P. A. M. Dirac, The principles of quantum mechanics, Oxford University Press, London (1958). 

\bibitem{Bell's theorem} J. S. Bell, Physics 1, 195 (1964).

\bibitem{Fine-Bell theorem} A. Fine, Phys. Rev. Lett. 48, 291 (1982).  
\bibitem{Cavalcanti non-commutativity and Bell inequality} E. G. Cavalcanti, C. J. Foster, M.D. Reid, and P. D. Drummond, Phys. Rev. Lett. 99, 210405 (2007).
\bibitem{Griffiths non-commutativity and Bell inequality}R. B. Griffiths, ``Quantum Nonlocality: Myth and Reality",
arXiv:1901.07050.

\bibitem{Aharonov weak value} Y. Aharonov, D. Z. Albert, and L. Vaidman, Phys. Rev. Lett. 60 (14), 1351 (1988).  

\bibitem{Lundeen complex weak value} J. S. Lundeen and K. J. Resch, Phys. Lett. A 334, 337 (2005).
\bibitem{Jozsa complex weak value} R. Jozsa, Phys. Rev. A  76, 044103 (2007). 

\bibitem{Johansen weak value as Bayesian efficient estimator} L. M. Johansen, Phys. Lett. A 322, 298 (2004).
\bibitem{Hall best estimate} M. J. W. Hall, Phys. Rev. A 69, 052113 (2004). 

\bibitem{Wiseman Bohmian mechanics} H. Wiseman, New J. Phys. 9, 165 (2007).

\bibitem{Ozawa error-disturbance UR} M. Ozawa, Phys. Rev. A 67, 042105 (2003).
\bibitem{Ozawa joint-measurement UR} M. Ozawa, Phys. Lett. A 320, 367 (2004).

\bibitem{Emerson violation of preparation independence} J. Emerson, D. Serbin, C. Sutherland, and V. Veitch, preprint at https://arxiv.org/abs/1312.1345 (2013).

\bibitem{Jaynes principle of indifference} E. T. Jaynes, Probability Theory: The Logic of Science, Cambridge University Press, Cambridge, 2003. 

\bibitem{Frauchiger-Renner no go theorem} D. Frauchiger and R. Renner, Nat. Comms.  9, 3711 (2018). 

\bibitem{Bohr classical language} N. Bohr, in P. A. Schilpp (ed.), Albert Einstein: Philosopher-Scientist, The Library of Living Philosophers, Volume 7, Open Court, Evanston (1949). 


\bibitem{Axioms for estimation error} A. Budiyono, in preparation. 
 
\end{thebibliography}
\end{document}